\documentclass[fleqn,usenatbib]{mnras}

\usepackage[T1]{fontenc}

\DeclareRobustCommand{\VAN}[3]{#2}
\let\VANthebibliography\thebibliography
\def\thebibliography{\DeclareRobustCommand{\VAN}[3]{##3}\VANthebibliography}

\usepackage{graphicx}	
\usepackage{amsmath}	
\usepackage{bm}
\usepackage{ulem}

\newcommand{\mtot}{M}
\newcommand{\rtot}{R}
\newcommand{\sigmasb}{\sigma_{\text{SB}}}
\newcommand{\rhobaryon}{\rho_{\text{B}}}
\newcommand{\m}{$^{-1}$}
\newcommand{\mm}{$^{-2}$}
\newcommand{\mmm}{$^{-3}$}
\newcommand{\ev}[1]{$#1$}

\title[Stationary accreted neutron star envelopes]{Stationary neutron star envelopes at high accretion rates}
\author[Nava-Callejas et al.]{
Mart\'in Nava-Callejas,$^{1}$\thanks{E-mail: mnava@astro.unam.mx (MNC)}
Yuri Cavecchi,$^{2}$
Dany Page$^{1}$
\\
$^{1}$Instituto de Astronom\'ia, Universidad Nacional Aut\'onoma de M\'exico, 
Apartado Postal 70-264, 04510 M\'exico, CDMX, Mexico\\
$^{2}$Departament de Fis\'{i}ca, EEBE, Universitat Polit\`ecnica de Catalunya, Av. Eduard Maristany 16, 08019 Barcelona, Spain
}

\date{Accepted XXX. Received YYY; in original form ZZZ}

\pubyear{2015}

\begin{document}
\label{firstpage}
\pagerange{\pageref{firstpage}--\pageref{lastpage}}
\maketitle

\begin{abstract}
In this work we model stationary neutron star envelopes at high accretion rates and describe our new code for such studies.
As a first step we put special emphasis on the rp-process which results in the synthesis of heavy elements and study in detail how this synthesis depends on the mass accretion rate and the chemical composition of the accreted matter.
We show that at very low accretion rate, $\dot{M} \sim 0.01 \dot{M}_{\text{Edd}}$, mostly low mass ($A\leq$ 24) elements are synthesized
with a few heavier ones below the $^{40}$Ca bottleneck.
However, once $\dot{M}$ is above ${\buildrel \sim \over >} 0.1 \dot{M}_{\text{Edd}}$ this bottleneck is surpassed and nuclei in the iron peak region ($A\sim$ 56) are abundantly produced.
At higher mass accretion rates progressively heavier nuclei are generated, reaching $A \sim 70$ at $\dot{M}_{\text{Edd}}$ and $A \sim 90$ at $5 \dot{M}_{\text{Edd}}$.
We find that when the rp-process is efficient, the nucleosynthesis it generates is independent of the accreted abundance of CNO elements as these are directly and copiously generated once the $3\alpha$-reaction is operating.
We also explore the efficiency of the rp-process under variations of the relative abundances of H and He.
Simultaneously, we put special emphasis on the density profiles of the energy generation rate particularly at high density beyond the hydrogen exhaustion point.
Our results are of importance for the study of neutron stars in systems in which X-ray bursts are absent but are also of relevance for other systems in describing the low density region, mostly below $10^6$ g cm\mmm, inbetween bursts.
\end{abstract}

\begin{keywords}
stars:neutron -- X-rays: bursts -- X-rays: binaries -- accretion
\end{keywords}



\section{Introduction}

Low-mass X-ray binaries (LMXBs) undergoing transient accretion episodes represent an astonishingly good opportunity to test evolution models of neutron stars, constraining properties such as 
surface gravity,  specific heat, thermal conductivity
and neutrino emission, as well as many details of the star's upper layers
\citep{2017JApA...38...49W}. 
This latter, low-density sector, commonly referred to as the \textit{envelope}, plays a crucial role in determining both the observed luminosity from the surface (visible after the outburst phase or during X-ray bursts) and 
inferring
the core's temperature during quiescence \citep{Shternin2007, Brown_2009, PhysRevLett.111.241102, PhysRevC.95.025806,Degenaar:2021aa}.

Gravitational energy released during accretion as heat at the surface is radiated away and has no impact on the internal temperature of the neutron star. 
However, compression of the crustal matter due to the increasing mass induces a series of nuclear reactions, such as electron captures, neutron emissions and pycnonuclear fusions \citep{Bisnovatyi-Kogan:1979aa,Sato:1979aa,1990A&A...227..431H} that have been dubbed as ``deep crustal heating'' by \citet{Brown:1998aa} (see, however, \citealt{Gusakov:2020aa,Gusakov:2021aa,Potekhin:2023aa}, for a different approach).
Most of the energy liberated by these processes, $\sim 1-2$ MeV per accreted nucleon, is released in the inner crust while a few electron capture reactions liberate a few tens of keV in the outer crust.
Once accretion stops, the surface of the neutron star is clearly visible and the star's cooling can be, and has been, directly observed. 
Detailed numerical modeling of this cooling phase has revealed the presence of another energy source, acting at lower densities and dubbed ``shallow heating'' \citep{Brown_2009}.
This source also releases up to a few MeVs of heat per accreted nucleon and has been found to act at densities below $10^9$ g cm$^{-3}$ in a few cases 
\citep{Degenaar:2014aa,Ootes:2016aa,Parikh:2018aa}
while in most cases it is acting deeper, up to $10^{10}$ g cm$^{-3}$ \citep{2015A&A...577A...5T,Vats:2018aa,Ootes:2019aa,Parikh:2019aa,Page_2022}
and even up to $10^{11}$ g cm$^{-3}$ \citep{2015A&A...577A...5T,Degenaar:2019aa}.

In the case of the first observed accretion outburst of the source MAXI J0556-332 the shallow heating may have been enormous \citep{Deibel_2015,Parikh:2017aa,Page_2022}, more than 10 MeVs per nucleon; however, it has been proposed that during this first outburst a giant ``hyperburst'' occurred in which case the shallow heating needed is then within the range deduced from the other sources \citep{Page_2022}.

The origin of this shallow heating has not yet been determined and it cannot be excluded that various different sources may be at work in different conditions as, e.g., accretion rate and/or state, star rotation rate (i.e. spin), outburst fluence, etc.
Various mechanisms have been proposed such as, e.g., conversion of gravitational energy into waves that deposit their energy much below the surface 
\citep{Inogamov:1999aa,inogamov2010}, electron captures and low-density pycnonuclear fusions \citep{Chamel:2020aa}.

Among transient sources, there exist a couple of remarkable cases whose luminosity, during the accretion outburst, is comparable to or even higher than the Eddington limit: XTE J1701-462 \citep{Homan:2006aa,Homan:2007aa,Lin:2009aa} and MAXI J0556–332 \citep{Homan:2011aa,Cornelisse:2012aa}.
Theoretical modeling suggests burning in accreting neutron star envelopes becomes stable at such high accretion rate \citep{1983ApJ...264..282P,1998ASIC..515..419B}
and observationally stable burning seems to begin even at a lower rate than predicted \citep{1998ASIC..515..419B,2021ASSL..461..209G}.
In the case of XTE J1701-462 no type I X-ray burst has been observed while it was accreting at high rate, but three short ones were observed at the end of its 2006-2007 outburst when the accretion rate had dropped to about 10\% of the Eddington luminosity \citep{Lin:2009ab}.
The situation is similar for MAXI J0556–332 in which case there is only one short type I burst observed, at the end of its 2020 outburst, when its accretion rate was also around 10\% of the Eddington limit \citep{Page_2022}.

According to numerical simulations, at accretion rates close to or above the Eddington luminosity the envelope reaches steady state H-burning via the rp-process (e.g. \cite{2001PhRvL..86.3471S, Fisker_2006, 2007ApJ...665.1311H}). In contrast to H-burning processes at lower accretion rates, such as the $pp$ and CNO cycles, a reliable simulation of the rp-process demands the inclusion of a large amount of nuclides and, consequently, of equations to be simultaneously solved \citep{1999ApJ...524.1014S,Fisker_2006}. 
Since this represents an important bottleneck for the simulation of the star evolution during high-accretion outburst episodes, devising a fast and accurate method for modeling this expensive, low-density portion of the star becomes imperative. 
A first step towards such scheme might be found in non-accreting systems, where only the core is explicitly evolved while the envelopes are approximated as being in steady-state and implemented as boundary conditions via $T_{b}$ - $T_{\text{eff}}$ relations, with $T_{\text{eff}}$ the effective temperature and $T_{b}$ the boundary temperature at the bottom of the envelope \citep{1983Gudmundsson, Potekhin:1997mn}. 
While the occurrence of type I bursts rules out the implementation of a similar scheme at all times during outburst episodes, a limited version can be devised exploiting the existence of inter-bursting periods and steady-states. 
As a first step towards the construction of such $T_{b}$ - $L_{b}$ - $T_{\text{eff}}$ relations, $L_{b}$ being the luminosity at the bottom of the envelope, the purpose of this paper is to present our stationary code for envelopes and apply it to explore the impact of changing physical parameters at high accretion rates. This, we anticipate, serves as both an update of the results from previous studies, a demonstration of the capabilities of our numerical code, 
and a preparation for future work.

The paper is organized as follows. 
In Section 2 we review the different timescales operating in the envelope and introduce its governing equations. 
Section 3 provides details on the numerical setup, boundary conditions and the network of reactions. 
In Section 4 the main results of this work are presented, and in Section 5 we provide a brief summary and prospects for future work.

\section{Stationary accreting neutron star envelopes}\label{sec:env_descript}

The envelope of a neutron star is defined as its outermost region occupying the layers of density $\rho\leq\rho_{b}$ where $\rho_{b}$, its boundary density, is usually taken between $10^{8}$ to $10^{10}$ g cm$^{-3}$. 
As such, it is very sensitive to temperature and composition, and the state of matter is expected to vary from an almost-ideal gas ($\rho\leq 10^{4}$ g cm$^{-3}$ and $T\geq 10^{5}$ K) to a gas of degenerate electrons imbued in a nuclide lattice (e.g. $\rho\geq 10^{8}$ g cm$^{-3}$ and $T\sim 10^{8}$ K). Due to the high gravitational acceleration at the surface of neutron stars, $g_{s}\sim G\mtot/\rtot^2\approx 10^{14}$ cm s$^{-2}$, the thickness of the envelope is 
very small, $d_{\text{Envelope}} \approx P/\rho g_{s}\approx 10$ m $\approx 10^{-3}\rtot$ (considering $P=P_b\sim 10^{25}$ erg cm$^{-3}$ at $\rho = \rho_b \approx 10^8$ g cm$^{-3}$ for a degenerate electron gas) where $\mtot$ and $\rtot$ are the star mass and radius, respectively, and $G$ the gravitational constant.
Assuming spherical symmetry for the whole star, we can infer the envelope mass to be $M_{\text{Envelope}} \approx 4\pi\rtot^{2}d_{\text{Envelope}}\rho_{b}\approx 10^{-9}M_{\odot}\approx 10^{-9}\mtot$.

The (time) evolution of the envelope is conditioned by the different timescales of several and simultaneous physical processes. Of relevance for the present paper are the following timescales:
\begin{itemize}
	\item The accretion timescale $\tau_{\text{acc}} = \delta M/\dot{M}$, expressing how much time it takes to accrete or replace a layer of mass $\delta M$ at a mass accretion rate $\dot{M}$. 
	If hydrostatic equilibrium holds, we have $\tau_{\text{acc}} = P/\dot{m}g_{s}$, where $\dot{m} = \dot{M}/4\pi R^2$ is the local mass accretion rate per unit area 
	and $P$ the pressure at the bottom of the accreted layer.
	For instance, the complete replacement of a typical neutron star envelope at Eddington accretion rate, i.e. $\dot{M}\sim 10^{18}$ g s$^{-1}$ (implying $\dot{m}\sim 10^{5}$ g cm$^{-2}$ s$^{-1}$ for a 10 km star), takes  $\tau_{\text{acc,b}}\approx 10^{25}/(10^{5}\times10^{14}) = 10^{6}$ s $\approx$ 12 days.         
	\item The nuclear timescale, denoting the characteristic time for energy release by nuclear reactions.
	At high temperatures, around and above $10^8$ K, hydrogen burns into He by the hot CNO cycle. 
	Its specific energy generation rate is $\dot{\varepsilon}_{\text{CNO}} = 4.6\times 10^{15} Z_\text{CNO}$ erg g\m s\m  \citep{2010ARNPS..60..381W} and it generates $\sim$6.8 MeV per nucleon or $E_h = 6.4\times 10^{18}$ erg g\m.
	We deduce the time to exhaust H as $\tau_\mathrm{nuc, CNO} = E_h/\dot{\varepsilon}_{\text{CNO}} = 1400/Z_\text{CNO}$ s which is about ten hours at solar abundances.
	The column depth at which H is exhausted is easily estimated as $y_\mathrm{nuc, CNO} = \dot{m} \tau_\mathrm{nuc, CNO}$.
	At the Eddington rate, one obtains $y_\mathrm{nuc, CNO} \approx 1.4\times 10^8/Z_\text{CNO}$ g cm\mm.
	The pressure at which this occurs is  $P_\mathrm{nuc, CNO}=g_s y_\mathrm{nuc, CNO}$ and the density, assuming a relativistic degenerate electron gas equation of state, is 
	$\rho_\mathrm{nuc, CNO} \approx 10^6$ g cm\mmm.
	\\
	In the case of the burning of He, the situation is more involved \citep{1998ASIC..515..419B}, but one can estimate the temperature and column density at which it starts, through the triple-$\alpha$ reaction: 
	$T_\mathrm{nuc, He} \approx 3.4\times 10^8 \dot{m}_5^{1/5}$ K and $y_\mathrm{nuc, He} \approx 5\times 10^7 \dot{m}_5^{-1/5}$ g cm\mm (the weak dependence on $\dot{m}$ is due to the high $T$ dependence of the triple-$\alpha$ rate). Then $\tau_\mathrm{nuc, He} \approx 5\times 10^2 \dot{m}_5^{-6/5}$ s.
	An important consequence of these estimates is that helium burning starts at a depth $y_\mathrm{nuc, He}$ that is lower than the one for H exhaustion, $y_\mathrm{nuc, CNO}$, and will thus produce C, O or N in the presence of H allowing for the occurrence of the rp-process.
	%

%
	\item The diffusion timescale, $\tau_{\text{diff}} = c_{P}\, d_{\text{Envelope}}^{2}\, K^{-1}$, determining how much time it takes for a thermal perturbation to dissipate if energy is carried away by radiation and conduction 	\citep{1969ApJ...156..549H}. Here, $c_{P}$ is the heat capacity per unit volume and $K$ the thermal conductivity. 
	At $\rho_{b} \sim 10^8$ g cm\mmm\ and $T_b \sim 10^8$ K, considering $K_{b}\sim 10^{16}$ erg cm\m s\m K\m  and $c_{P,b}\sim 10^{14}$ erg cm\mmm K\m \citep{PhysRevLett.111.241102}, we have $\tau_{\text{dif},b}\approx 10^{4}$ s $\approx$ a few hours. 
	At $\rho_\mathrm{CNO}$, however, it is only $\tau_\text{diff,CNO} \approx 10^{2}-10^3$~s $\approx$ several minutes.
\end{itemize}
We thus see that under these conditions $\tau_\mathrm{nuc, CNO} \approx \tau_\mathrm{acc,CNO} \gg \tau_\mathrm{diff, CNO}$ meaning that the heat released by nuclear reactions has ample time to diffuse away while the reverse inequalities are associated with thermonuclear instabilities \citep{1979ApJ...233..327T, 1980ApJ...241..358T}.

For isolated neutron stars, it is frequently assumed that the envelope temperature and luminosity are stationary since, at the linear order, perturbations decay on a timescale $\tau_{\text{diff}}$, short with respect to the timescales of interest (see for instance \citealt{1969ApJ...156..549H, 1983Gudmundsson, Potekhin:1997mn}). 
Albeit useful, this assumption breaks down in the accreting scenario since this argument neglects the effects from perturbing heating sources. However, considering the success and limitations of previous works in understanding certain characteristics of the envelope during accretion periods applying stationary models \citep{1999ApJ...524.1014S, 2000ApJ...544..453C, 2003ApJ...599..419N}, as demonstrated by the good agreement with fully time-dependent numerical simulations \citep{Fisker_2006, 2007ApJ...665.1311H, 2014MNRAS.445.3278Z}, properly defined stationary states have a physical meaning we can use to describe specific intervals of the neutron star evolution. Consequently, as long as we restrain from resolving type I bursts, stationary solutions to the envelope equations are useful either as coupled with numerical codes following the thermal evolution of the rest of the neutron star, or as initial conditions for numerical codes simulating the evolution of the envelope itself, e.g., \texttt{MESA} \citep{Paxton_2011, Paxton_2015, Meisel_2018} or \texttt{KEPLER} \citep{Weaver_1978, 2023arXiv230510627T}.

\subsection{Equations governing the envelope}\label{subsec:odes}

A full description of a neutron star interior is obtained by solving two sets of nonlinear partial differential equations, the ones of \textit{structure} and \textit{thermal evolution}, extensively discussed elsewhere (e.g. \citealt{1975PASJ...27..197S,1977ApJ...212..825T,1979AnPhy.118..341I,Shapiro:1983wz,Townsley_2004,2011RSPSA.467..738L,10.1093/mnras/sty1725}). 
In our stationary approximation, they reduce to nonlinear ordinary differential equations. Since pressure is a continuous variable with prominent variations (e.g. $P_{b}\sim 10^{25}$ erg cm$^{-3}$ to $P_{s}\sim 10^{14}$ erg cm$^{-3}$ against $\rtot - r_{b}\approx d_{\text{Envelope}}\sim 10$ m), it is best to employ $P$ instead of $r$ as the independent variable. Considering a nonzero redshifted mass accretion rate $\dot{M}^{\infty}\neq 0$, the equation describing our system are
\begin{align}
\frac{dr}{dP} & = -\frac{1}{g\, \mathcal{H}\mathcal{G}e^{\Lambda}}, \label{eq:ss_eq1set}\\
\frac{dm}{dP} & = 4\pi r^{2}\rho\frac{dr}{dP}, \label{eq:ss_eq2set}\\
\frac{dY_{i}}{dP} & = -\frac{4\pi r^{2}\rhobaryon e^{\Phi + \Lambda}}{\dot{M}^{\infty}}\frac{dr}{dP}\mathcal{R}_{i}\ \ \forall i\in\left\{1,\ldots,N_{\text{nuclides}}\right\}, \label{eq:ss_eq3set}\\
\frac{dT}{dP} & = \frac{T}{P}\nabla_{T},\label{eq:ss_eq4set}\\
\nabla_{T} & := -\frac{3\kappa\rhobaryon LPe^{\Lambda}}{64\pi\sigmasb T^{4}r^{2}}\frac{dr}{dP} + \left(1 - \frac{\rho}{\mathcal{H}}\right), \nonumber\\
\frac{dL}{dP} & = \frac{2L}{c^{2}\mathcal{H}} + 4\pi r^{2}\rhobaryon e^{\Lambda}\frac{dr}{dP}\left(\dot{\varepsilon}_{\text{nuc}} + \dot{\varepsilon}_{\text{grav}} - \dot{\varepsilon}_{\nu}\right) \label{eq:ss_eq5set}
\end{align}
where, following \cite{1977ApJ...212..825T},
\begin{align}
g(r,m) & = \frac{Gm}{r^{2}}e^{\Lambda(r)} = \frac{Gm}{r^{2}}\left(1 - \frac{2Gm}{c^{2}r}\right)^{-1/2}, \label{eq:thorne_1} \\
\mathcal{H} = \rho + \frac{P}{c^{2}} \ \ ;&\ \mathcal{G} = 1 + \frac{4\pi Pr^{3}}{mc^{2}}.\label{eq:thorne_2}
\end{align}
Here, $(m,T,\kappa, L, \rhobaryon)$ denote the gravitational mass, temperature, opacity, luminosity and rest-mass density; $\rho$ is the total density, i.e. $= \rhobaryon + $ Internal energy times $c^{2}$, while ($Y_{i}$, $\mathcal{R}_{i}$) denote the abundance and the total creation/annihilation rate of the i-th species in the mixture. By $\dot{\varepsilon}_{j}$ we denote the specific energy generation rate due to the j-th process, and $\sigmasb$ denotes the Stefan-Boltzmann constant. Finally, we denote $g_{s} = g(\rtot,\mtot)$ as the \textit{local surface gravity}. A few more details concerning these equations are described in Appendix \ref{sec:odes_extended}.

\section{Methodology}\label{sec:methods}

\subsection{Numerical implementation and boundary conditions}\label{subsec:boundary_con}

Considering the stiff character of Eqs.~\eqref{eq:ss_eq3set} \citep{1986A&A...162..103M, 2006NuPhA.777..188H}, for the numerical solution of Eqs.~\eqref{eq:ss_eq1set} - \eqref{eq:ss_eq5set} we employed the variable-order Bader-Deuflhard method as implemented by \cite{Press:1992vz}. 
Equations are solved in in the range $[P_{s}, P_{b}]$, with $P_b$ the pressure associated with $\rho_b$ and $P_s$ with the surface. At $P_{s}$ we impose $m_{s} = \mtot$, $r_{s} = \rtot$ and the Schwarzschild condition $\Phi(\rtot) = -\Lambda(\rtot)$. 
Since we are only interested in modeling the emission from the star surface, we parameterize luminosity as $L_{s} = 4\pi \rtot^{2}\sigmasb T^{4}_{\text{eff}}$ with $T_{\text{eff}}$ the effective temperature. ($P_{s}$, $T_{s}$) are computed via the standard Eddington relations,
\begin{align}
T^{4}_{s} & = \frac{3}{4}\left(\tau_{s} + \frac{2}{3}\right)T^{4}_{\text{eff}}\\
P_{s} & = \frac{\tau_{s}g_{s}}{\kappa_{s}} + \frac{4\sigma_{\text{SB}}}{3c}T^{4}_{s}.
\end{align}
at optical depth $\tau_{s} = 2/3$. For models with Sun-like composition at the surface, we adopt the abundances from \cite{1989GeCoA..53..197A}. 
Further details on the equation of state (EOS), opacity and thermal neutrinos can be found in Appendix \ref{sec:microphysics}. 
For the mass accretion rate unit, we adopt Eddington's one considering a hydrogen mass fraction $X = \text{0.71}$ and a 10 km radius object, $\dot{M}_{\text{Edd}} = $ 1.1 $\times 10^{18}$ g s$^{-1}$. 
When quoting a mass accretion rate we measure it ``at infinity'', $\dot{M}^\infty$, and the corresponding value measured by an observer within the star will be $\dot{M}  =  e^{-\Phi}\dot{M}^{\infty}$.
We discuss luminosity in either c.g.s. or MeV baryon$^{-1}$ units, related via $Q_{L} = m_{u} L / \dot{M} = e^{-\Phi} m_{u} L^\infty / \dot{M}^\infty = e^{-\Phi} Q_{L}^\infty$, with $m_{u}$ the atomic mass unit. 
(An observer at infinity will measure a luminosity $L^\infty = e^{2\Phi} L$.)
For comparison, taking $\dot{M} = \dot{M}_{\text{Edd}}$ and $L = 10^{37}$ erg s$^{-1}$, $Q_{L}\approx 9.42$ MeV baryon$^{-1}$.

\subsection{Network of Nuclear Reactions}\label{subsec:nucnet}

Around and above $\dot{M}_{\text{Edd}}$, simple networks burning H and He (such as those in \citealt{2000ApJ...544..453C, 2003ApJ...599..419N}) cease to be useful since the rp-process requires hundreds of nuclides and reactions in order to (i) obtain the correct $\dot{\varepsilon}_{\text{nuc}}$; (ii) guarantee the complete depletion of hydrogen at high densities \citep{1996ApJ...459..271T, 1997ApJ...484..412R, 1999ApJ...524.1014S} and (iii) include the majority of isotopes in the valley of stability \citep{1981ApJS...45..389W, 2001PhRvL..86.3471S}. While the \texttt{rp\_298} network from \cite{Fisker_2006} satisfies (i) and (ii), it omits a considerable amount of isotopes above $A=64$ in the valley of stability. Therefore, we extended this network up to 380 nuclides, sufficient to explore the physics of stationary states below 5$\dot{M}_{\text{Edd}}$ (see the full list in Table~\ref{tab_network_net}). All binding energies per baryon $\mathcal{B}_{j}/A_{j}$ were taken from \cite{nudat3}. 

We included all $(p,\gamma)-(\gamma,p)$, $(\alpha,\gamma)-(\gamma,\alpha)$, $(\alpha,p)-(p,\alpha)$ thermonuclear reactions, as well as those for the $pp$ I-IV chains, $3\alpha\to\, ^{12}$C, carbon and oxygen burning, adopting the recommended versions of the fits from the \citet[][see also \citealt{Cyburt_2010}]{jinareaclib}. Whenever necessary to refer to a specific version of a rate, we use the assigned labels in the JINA Library, such as \texttt{rath} or \texttt{il10} below. Corrections due to screening were accounted for following \cite{1973ApJ...181..457G, 1978ApJ...226.1034A, 1979ApJ...234.1079I}\footnote{We adopted the implementation of screening from \url{https://cococubed.com/code_pages/burn.shtml}}. 

For weak rates we employed the $\rho, T$-independent versions of $\beta^{+}$ decays. 
This approximation is adequate since (i) at the range of interest electron captures are less frequent than experimentally-constrained $\beta^{+}$ decays and (ii) the coarse character of existing tables compromises the accuracy of mass fractions profiles depending on the employed interpolation scheme \citep{1985ApJ...293....1F, 2013PhRvC..88a5806T, Paxton_2015}. 
Regarding electron captures, the importance of the $A = 56$ group motivated us to incorporate the fits over $^{56}$Ni and $^{56}$Co, as detailed in Appendix \ref{apx_sec:ec_fits}. 
For the neutrino energy losses from these reactions, $\dot{\varepsilon}_{\nu}$, we consider the treatment from \citet{1975ARA&A..13...69F} and \citet{2006NuPhA.777..188H}.

\begin{table}
	\centering
	\caption{List of nuclides in the network.}
	\begin{tabular}{|p{0.5cm}|p{0.90cm}|p{0.5cm}|p{1.1cm}|p{0.5cm}|p{1.0cm}|}
		\hline
		$Z$ & $A$ & $Z$ & $A$ & $Z$ & $A$\\
		\hline
		\hline
		H & 1-2 & K & 35-39 & Rb & 74-85\\
		He & 3-4 & Ca & 36-44 & Sr & 75-88\\
		Li & 6-7 & Sc & 39-45 & Y & 78-89\\
		Be & 7,9 & Ti & 40-50 & Zr & 79-90\\
		B & 8,10,11 & V & 43-51 & Nb & 82-90\\
		C & 11-13 & Cr & 44-54 & Mo & 83-90\\
		N & 12-15 & Mn & 47-55 & Tc & 86-90\\
		O & 13-18 & Fe & 48-58 & Ru & 87-91\\
		F & 17-19 & Co & 51-59 & Rh & 89-93\\
		Ne & 18-22 & Ni & 52-62 & Pd & 90-94\\
		Na & 20-23 & Cu & 56-63 & Ag & 94-98\\
		Mg & 21-26 & Zn & 57-68 & Cd & 95-99\\
		Al & 22-27 & Ga & 60-69 & In & 98-104\\
		Si & 24-30 & Ge & 61-74 & Sn & 99-105\\
		P & 26-31 & As & 65-75 & Sb & 106\\
        S & 27-34 & Se & 66-78 & Tc & 107\\
        Cl & 30-35 & Br & 70-79 & & \\
        Ar & 31-38 & Kr & 71-84 & & \\
		\hline
	\end{tabular}\label{tab_network_net}
\end{table}

\begin{figure*}
\includegraphics[width=0.49\linewidth]{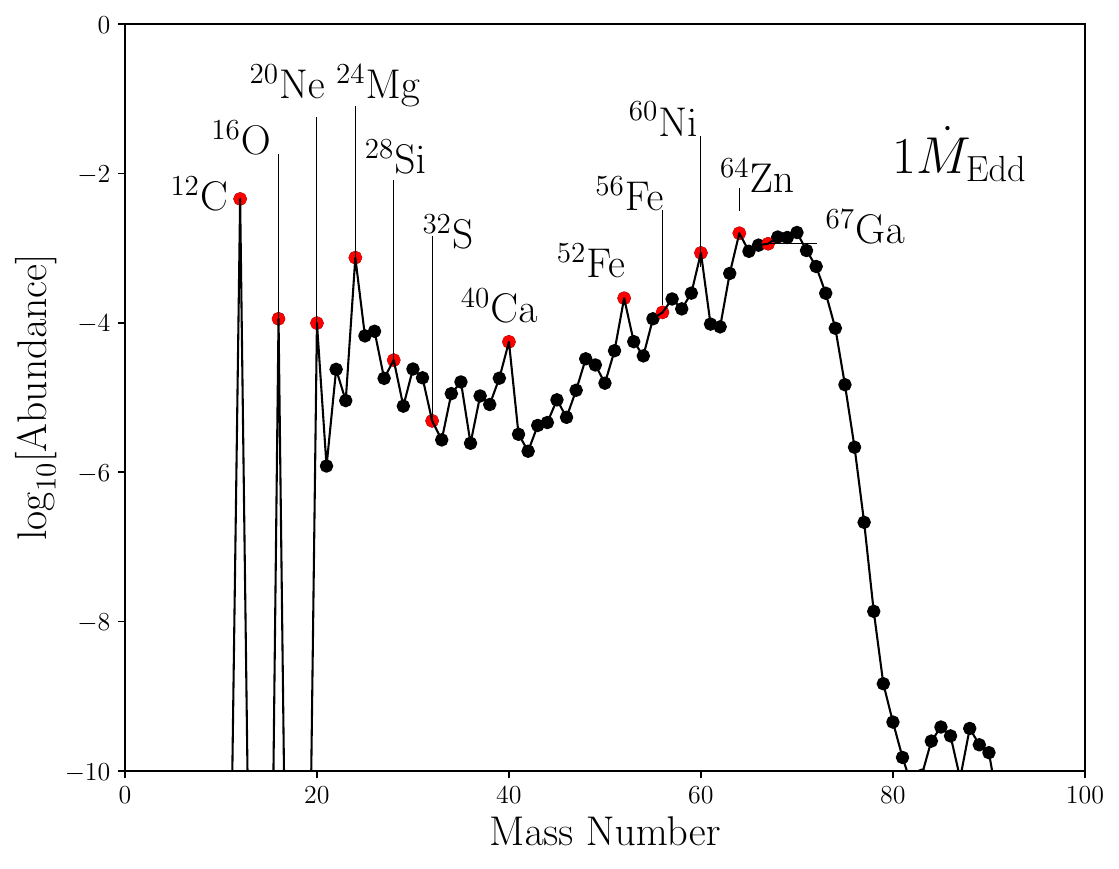}
\hspace{\stretch{1}}
\includegraphics[width=0.49\linewidth]{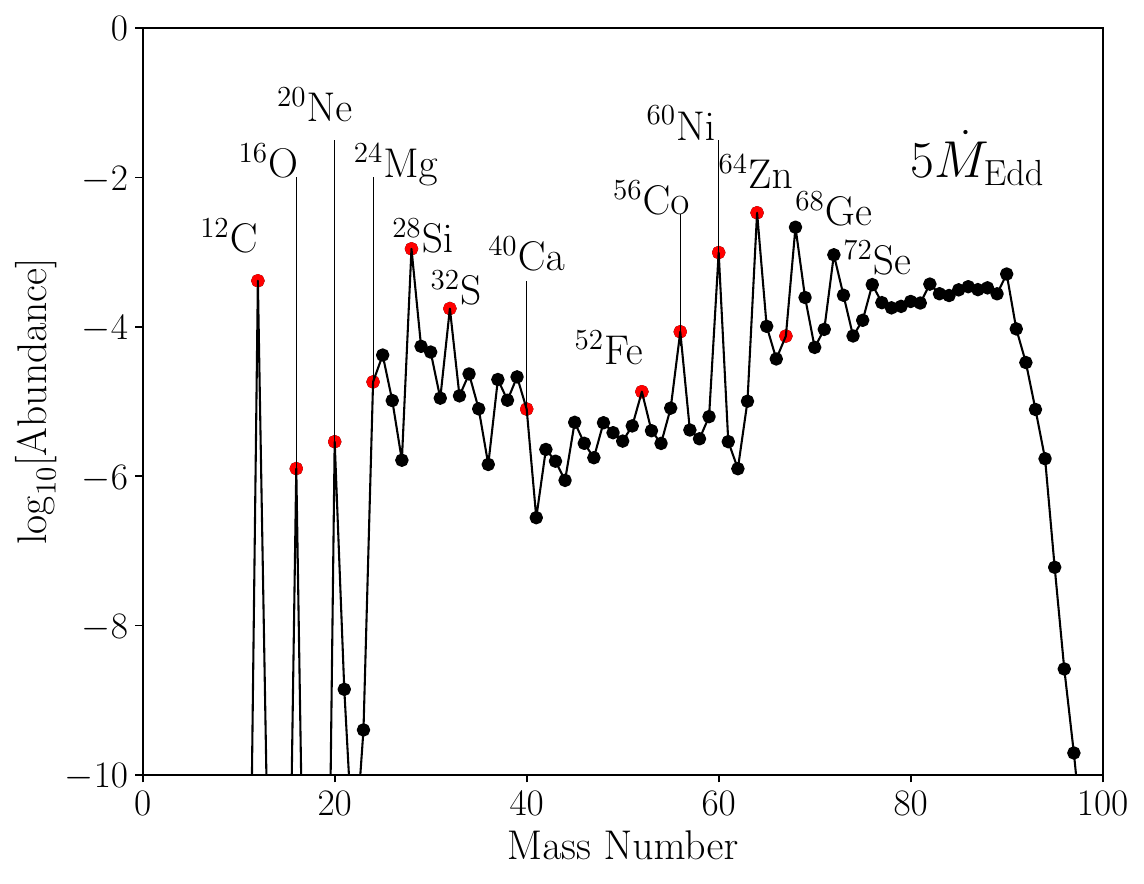}
 \caption{Distribution of abundances at $10^{7}$ g cm$^{-3}$ for $\dot{M}^{\infty} = \dot{M}_{\text{Edd}}$ (left) and $5 \dot{M}_{\text{Edd}}$ (right). 
 For some selected isobars, red dots, we indicate the most abundant nucleus.}
\label{fig:abun_1_5mdot}
\end{figure*}

\begin{figure*}
\centering
\includegraphics[width=\linewidth]{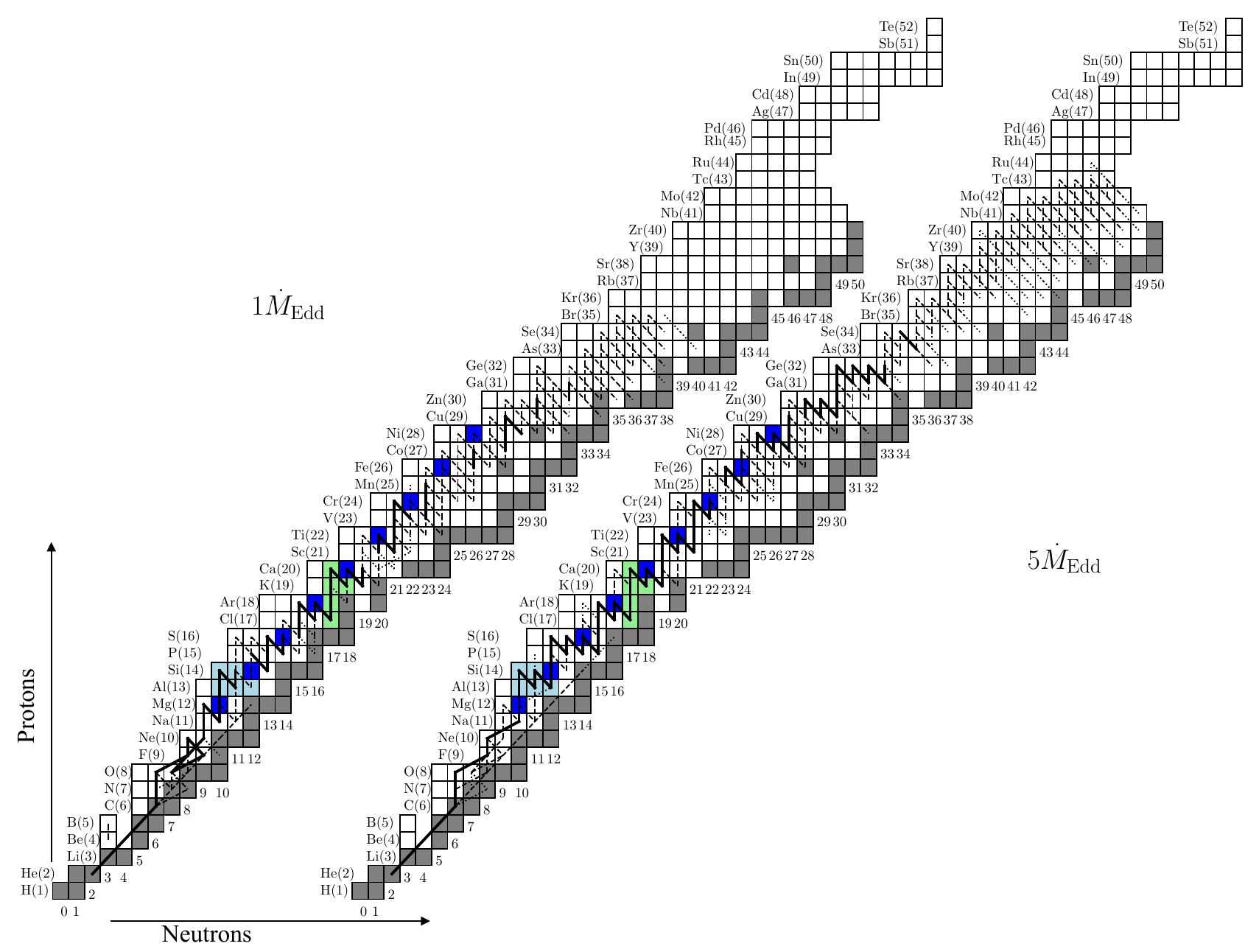}
\caption{Space-integrated reaction flows for $\dot{M}^\infty = \dot{M}_{\text{Edd}}$ and $\dot{M}_{\text{Edd}}$, in units of the corresponding value of the $3\alpha$ reaction. 
Solid lines: $\geq$ 0.3; dashed lines: between 0.3 and 0.01; dotted lines correspond to values between 0.01 and 0.005. 
Nuclides colored in lightblue illustrate the ``sawtooth'' path, while those in green correspond to the $\beta-3p-\beta$ path (see main text for further details). 
Some $T_{z} = -1$ nuclides are colored in blue.}
\label{fig:nucchart_1mdot}
\end{figure*}

\section{Results}\label{sec:results}

\subsection{Envelopes at 1 and 5 times Eddington}\label{subsec:res_medd}

One of the most robust networks for accreting neutron stars above $\dot{M}_{\text{Edd}}$ was presented by \cite{1999ApJ...524.1014S}. 
Therefore, as a solid test for our numerical code, we first compare our results against theirs for $\dot{M}^\infty = \dot{M}_{\text{Edd}}$ and $\dot{M}^\infty = 5\dot{M}_{\text{Edd}}$, fixing $\rho_{b} = 10^{7}$ g cm$^{-3}$, setting Solar composition at the surface and selecting appropriate $T_{\text{eff}}$ allowing us to obtain $T_{b}\approx \text{5.45}\times 10^{8}$ K for $\dot{M}_{\text{Edd}}$ and $8\times 10^{8}$ K for $5\dot{M}_{\text{Edd}}$, as suggested by their Fig. 5.

The resulting distributions of abundances at $\rho_b$ are shown in the left and right panels of Fig.~\ref{fig:abun_1_5mdot}, and the integrated reaction flows (see Eq.~\ref{eq:Fij}) are plotted in Fig.~\ref{fig:nucchart_1mdot}. In general, we find a very good agreement. 
For instance, at $\dot{M}_{\text{Edd}}$ our code reproduces the double-peaked distribution, corresponding to $^{12}$C, $^{16}$O, $^{20}$Ne, and $^{24}$Mg at low A's and the $rp$-synthesized $^{64}$Zn - $^{67}$Ga on the other side. 
In the corresponding integrated reaction flow we also observe the competition between the \textit{sawtooth} path (illustrated by the lightblue-colored nuclides between $^{22}$Mg and $^{26}$Si) and the $\beta-3p-\beta$ path (as exemplified by the lightgreen-colored nuclides between $^{34}$Ar and $^{38}$Ca) among $T_{z} = -1$ nuclides, where $T_{z}$ is the 3rd projection of the isospin, i.e. half the difference between a nuclide neutron and proton numbers \citep{Fisker_2006}. 
At $5\dot{M}_{\text{Edd}}$, we reproduce local maxima in the distribution such as those at $^{24}$Mg, $^{28}$Si, $^{64}$Zn, $^{68}$Ge and $^{72}$Se. The discrepancies between our results and those from \cite{1999ApJ...524.1014S} can be explained as follows:
\begin{itemize}
	\item \textbf{Absence of low-charge nuclides.} In our integrated flow we obtain a strengthening in the $^{7}$Be($p,\gamma$)$^{8}$B / $3\alpha$ ratio. Since the distribution of final abundances below $A=12$ does not exhibit additional peaks, we infer such strengthening is an artifact of our network since we do not include $^{8}$Be, $^{9}$B, $^{9} - ^{10}$C and their connecting reactions with the rest of the isotopes.
	\item \textbf{Different versions of thermonuclear reaction rates.} For instance, the $^{19}$Ne($\beta^{+}\nu_{e}$)$^{19}$F($p,\alpha$)$^{16}$O path in the integrated flow occurs because the recent $^{19}$Ne($p,\gamma$)$^{20}$Na recommended rate, \texttt{il10}, is up to $10^{5}$ times smaller than the old \texttt{rath} version. 
	Another source of discrepancy is the absence of local maxima at $A = 44$ and $A = 56$ in the abundance distribution of ashes, which can be attributed to the employed versions of $^{44}$Ti($p,\gamma$)$^{45}$V (\texttt{nfis}, $10^{1}\ -\ 10^{2}$ times larger below $10^{8.5}$ K than \texttt{laur} and \texttt{rath}) and $^{56}$Ni($p,\gamma$)$^{57}$Cu rates, \texttt{wien(v1)}, which has an important impact on the final abundance of $^{56}$Ni, \citep[see e.g.][]{PhysRevC.64.045801}. Other thermonuclear rates we identify as responsible for altering the integrated flow and having some impact on the final abundances at $\dot{M}_{\text{Edd}}$ are $^{28}$Si($p,\gamma$)$^{29}$P, $^{39}$Ca($p,\gamma$)$^{40}$Sc and $^{45}$V($p,\gamma$)$^{46}$Cr.
	\item \textbf{Updated binding energies.} Paths such as $^{21}$Mg($p,\gamma$)$^{22}$Al or $^{64}$Ge($p,\gamma$)$^{65}$As are less favorable since their $Q$-value is now negative, -0.892 and -0.090 MeV respectively. The first blocking inhibits the enhancement of $^{23}$Al via $^{22}$Al ($p,\gamma$) $^{23}$Si ($\beta^{+}\nu_{e}$) $^{23}$Al and conditions the whole evolution to the competition between $^{21}$Na ($\beta^{+}\nu_{e}$) $^{21}$Ne and $^{21}$Na ($p,\gamma$) $^{22}$Mg rates. Also, on one hand, the second blocking only contributes to a small decreasing(enhancing) of the final $A=64$(65) abundances, while, on the other hand, $^{25}$Si($p,\gamma$)$^{26}$P is now viable since $Q = $0.140 MeV and thus favours the $^{26}$P ($p,\gamma$) $^{27}$S ($\beta^{+}\nu_{e}$) $^{27}$P path over $^{25}$Si ($\beta^{+}\nu_{e}$) $^{25}$Al ($p,\gamma$) $^{26}$Si ($p,\gamma$) $^{27}$P.
	\item \textbf{Impact of weak rates.} While the integrated flow and the final abundances are not significantly different, we infer these rates might enhance the available metals at lower densities, the synthesis of additional $A = 80$ - $100$ nuclides and the most abundant isotope at each $A$. For instance, by employing the electron capture fits for $A=56$ isotopes we obtained a high(low) $^{56}$Fe($^{56}$Ni) abundance in the $\dot{M}_{\text{Edd}}$ case since the rate for $\beta^{+}$-decay from $^{56}$Ni is hundred-to-thousand times smaller than the electron capture rate.
\end{itemize}
%

\begin{figure*}
\includegraphics[width=0.99\linewidth]{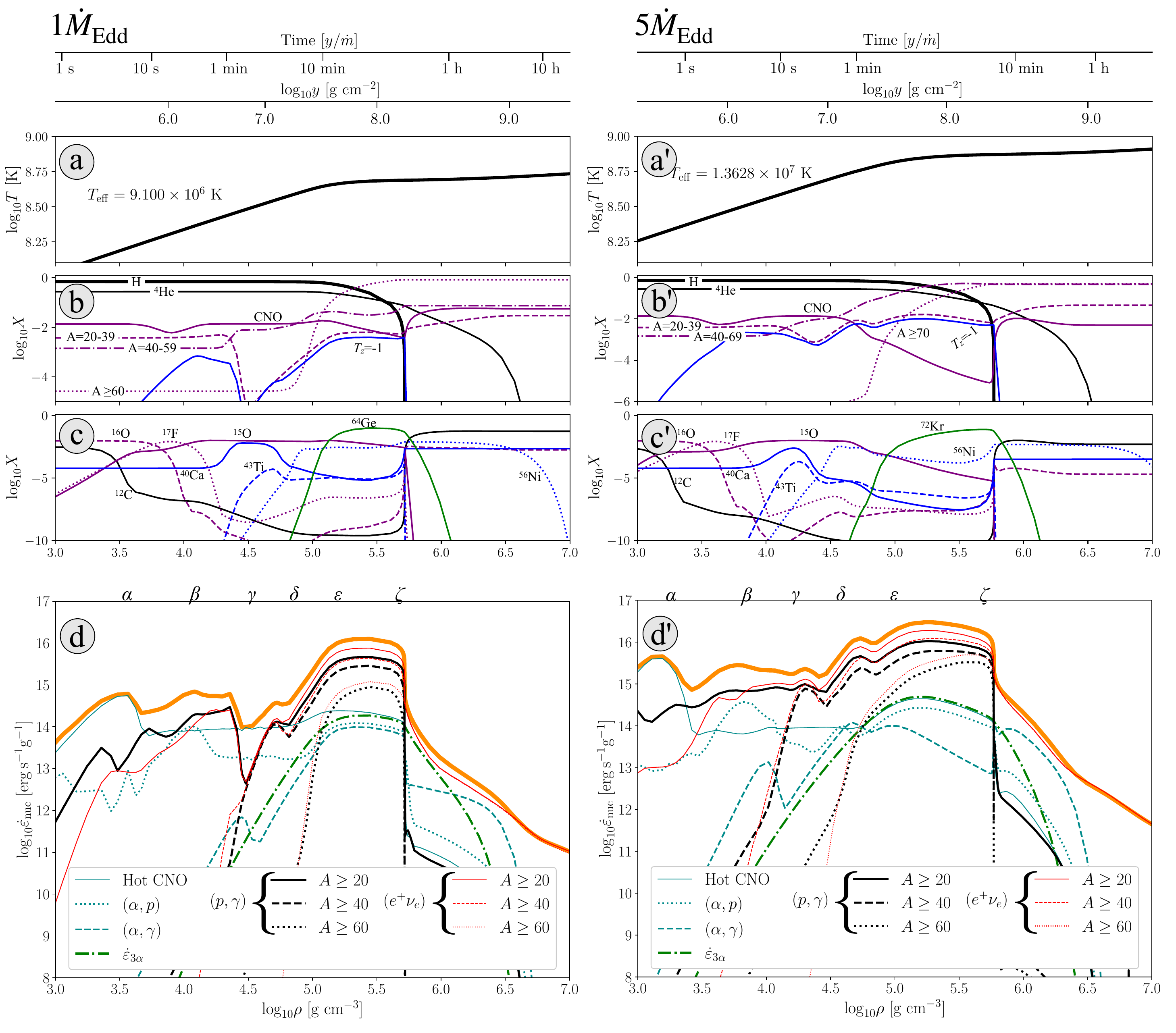}
\caption{Envelope models at $\dot{M}^{\infty} = \dot{M}_{\text{Edd}}$ (left panels) and $5\dot{M}_{\text{Edd}}$ (right panels) as a function of $\rho$.
The upper scales indicate the corresponding column depth $y$ and time spent by the accreted matter since it started its journey from the neutron star surface.
Panels (a) and (a'): temperature;
panels (b) and (b'): mass fractions of selected light nuclei; 
panels (c) and (c'): mass fractions of selected heavy nuclei; 
panels (d) and (d'): specific energy generation of dominant processes, as indicated, and with the upper thick line showing the total energy generation.
Greek labels at the top of panels (d) and (d') indicate specific positions of events discussed in the text.
}
 \label{fig:1_5medd_tx1}
\end{figure*}

We now describe the main features of these models as displayed in Fig.~\ref{fig:1_5medd_tx1} where we present the temperature profiles, mass fractions of selected interesting nuclei and details on the energy generation due to nuclear reactions in the $\text{log}_{10}\rho$ range 3 - 7. 
As discussed in Appendix \ref{sec:odes_extended} (specifically, the discussion around Eq.~\ref{eq:egravenuc1}), nuclear reactions play a small role in the density range $\text{log}_{10}\rho\leq 3.0$ and are thus of little interest for the present discussion.
To facilitate reference, all discussed events are noted at the top of panels (d) and (d') by Greek letters.

First, consider event \ev{\alpha} in panels (d) and (d'): the energy generation is dominated by the burning of the accreted carbon through $^{12}$C($p,\gamma$)$^{13}$N($p, \gamma$)$^{14}$O, which takes only a few seconds.
Exhaustion of $^{12}$C is clearly seen in panels (c) and (c').
On longer timescales, up to event \ev{\delta}, the hot CNO is fully operating at an almost constant rate, consequence of the $^{14}$O($\beta^+ \nu_e$)$^{14}$N and $^{15}$O($\beta^+ \nu_e$)$^{15}$N decays with respective half lives of 70 and 122 sec.
Between events \ev{\delta} and \ev{\varepsilon} an enhancement of the rate of hot CNO takes place as a consequence of the $3\alpha$-reaction, which, due to $\text{log}_{10}T\sim$ 8.5, injects more  $^{12}$C into the CNO cycle where carbon is then
almost immediately converted into $^{14}$O and $^{15}$O.
In the $\dot{M} = \dot{M}_{\text{Edd}}$ envelope this shows itself as an increase in the total CNO nuclei mass fraction while in the $\dot{M} = 5 \dot{M}_{\text{Edd}}$ case temperature is so high that in spite of this $^{12}$C injection CNO nuclei are consumed faster than they are produced and their total mass fractions decreases.
At event \ev{\zeta}, the $^{1}$H exhaustion point, the hot CNO cycle naturally shuts off.

At densities in the range of $10^{3.5} - 10^{4.5}$ g cm\mmm, event \ev{\beta}, we see a dramatic attempt at a CNO breakout:
there is rapid consumption of $^{16}$O from $^{16}$O($p,\gamma$)$^{17}$F, resulting in a significant decrease of the total CNO nuclei mass fraction, seen in panels (b) and (b').
However, this is then followed by the path $^{17}$F($p,\gamma$)$^{18}$Ne($\beta^+ \nu_e$)$^{18}$F($p,\alpha$)$^{15}$O which returns the escaped $^{17}$F back to the CNO nucleus $^{15}$O, having the net effect of converting the available $^{16}$O into $^{15}$O, leaving almost unaltered the total mass fraction of CNO species and thus resulting in an extension of the CNO cycle into a CNOFNe cycle. This can be seen in the resulting evolutions of the $^{15-16}$O and $^{17}$F mass fractions, panels (c) and (c'), as well as at the local maximum in the ``($\alpha,p$)'' energy generation curve (which includes the inverse reactions ($p,\alpha$) as well), panels (d) and (d').

Once the initial $^{12}$C has been burnt, event \ev{\alpha}, energy generation is dominated by the ($p,\gamma$) proton captures, further dubbed as $p$-captures, over nuclei in the $A$ range 20 - 39, and at slightly higher densities on nuclei with $A\geq 40$, as illustrated in panels (d) and (d'). 
For nuclei with $Z$ larger than 8, $\alpha$-captures become very limited by Coulomb repulsion so these processes become important only at later stages, between events \ev{\delta} and \ev{\zeta}, when the temperature has significantly risen.
The early occurrence of $20\leq A\leq 39$ $p$-captures is a consequence of, mainly, two factors: first, some $^{14}$O($\alpha,p$)$^{17}$F and $^{15}$O($\alpha,\gamma$)$^{19}$Ne CNO breakout reactions allow the buildup of $^{19-20}$Ne and $^{20-21}$Na material. Next, $p$-captures over these nuclides can proceed enhancing the amount of proton-rich isotopes whose half-lives against $\beta^{+}$ decays are typically shorter than $10$ s, facilitating the occurrence of further $(\beta^{+}\nu_{e})$ - $p$-capture chains. In this regard, the first stage of the rp-process starts operating.
Due to the importance of $(p,\gamma)$ and $\beta^{+}$-decay reactions in the integrated flow, we explicitly plot their respective contributions in panels (d) and (d'), considering three different nuclide mass ranges. Note that, in contrast to the hot CNO cycles, where just a few reactions can be identified as the source of $\geq 90\%$ of the total energy, the released energy in the rp-process comes from the whole collection of reactions instead of just a few ones.

Let us now consider the buildup of nuclei above $A=40$.
At $\dot{M}_{\text{Edd}}$, the $^{43}$Ti($\beta^{+}\nu_{e}$)$^{43}$Sc($p,\alpha$)$^{40}$Ca path slows down the buildup of $^{43}$Ti, but results in a tremendous enhancement of $^{40}$Ca,
event \ev{\gamma}.
Near $\text{log}_{10}\rho = 5.0$, sufficient $^{43}$Ti and temperature have been reached as to allow $^{43}$Ti($p,\gamma$)$^{44}$V, allowing material to reach $A>43$ via subsequent $p$-captures and $\beta^{+}$ decays. 
Such bottleneck seems less accentuated at $5\dot{M}_{\text{Edd}}$, where the $^{40}$Ca and $^{43}$Ti buildups take place at lower densities in comparison, $\text{log}_{10}\rho = 4.5$. 
This allows the reaction flow to reach up to $^{56}$Ni below $\text{log}_{10}\rho = 4.75$ at $5\dot{M}_{\text{Edd}}$, while we must await until $\text{log}_{10}\rho \sim 5.25$ at $\dot{M}_{\text{Edd}}$, 
events \ev{\delta} and \ev{\varepsilon} in the energetics diagram. 

At the next events, from \ev{\varepsilon} to \ev{\zeta}, we see a third stage of the rp-process, enhancing the abundance of proton-rich nuclides such as $^{56}$Ni, $^{64}$Ge at $\dot{M}_{\text{Edd}}$ and $^{72}$Kr at $5\dot{M}_{\text{Edd}}$. 
The generated energy in both scenarios reaches a global maximum, of $10^{16}$ erg g$^{-1}$ s$^{-1}$ in the first case and almost ten times larger in the second. 
Despite this enhancement of $A > 60$ material, as well as the considerable contribution from the $3\alpha$-process, comparable with that from the typical hot CNO, the released energy mostly comes from $p$-captures over $A\leq 40$ material, as well as the subsequent $\beta^{+}$ decays.
At event \ev{\zeta}, $\text{log}_{10}\rho \simeq 6$, the available hydrogen has significantly decreased ($X_{s}\sim$ 0.70 $\to$ $X_{\rho = 10^{6}} \sim 10^{-10} $) resulting in a sudden drop of all proton capture reactions at this density in both models.

After this event, the fraction of proton-rich material begins to drop as a consequence of $\beta^{+}$ decays, which dominate the released energy up to $\rho_{b}$.
This is illustrated through the examples of two nuclei far away from the stability line, $^{64}$Ge at $\dot{M}_{\text{Edd}}$ and  $^{72}$Kr at $\dot{M}_{\text{Edd}}$, in panels (c) and (c').
For $5\dot{M}_{\text{Edd}}$, hydrogen exhaustion led to the synthesis of $A\geq 70$ material. 
The energy release is now dominated by $\beta^{+}$ decays of $A\leq 60$ isotopes towards the valley of stability at $\dot{M}_{\text{Edd}}$ and $A\leq 40$ at $5\dot{M}_{\text{Edd}}$,
with still a significant contribution from the $3\alpha$ and $(\alpha, \gamma)$ reaction families.
Since the $A = 20$ - $60$ abundances around $\rho_b$ are of the same order of magnitude, $\sim 10^{-5}$ according to Fig.~\ref{fig:abun_1_5mdot}, the amount of released energy is very similar at both accretion rates, and explains the small $T$ increase in both cases, $\text{log}_{10}T = $ 8.66 to 8.69 at $\dot{M}_{\text{Edd}}$ and from 8.83 to 8.85 at $5\dot{M}_{\text{Edd}}$.

\begin{figure}
\includegraphics[width=0.99\linewidth]{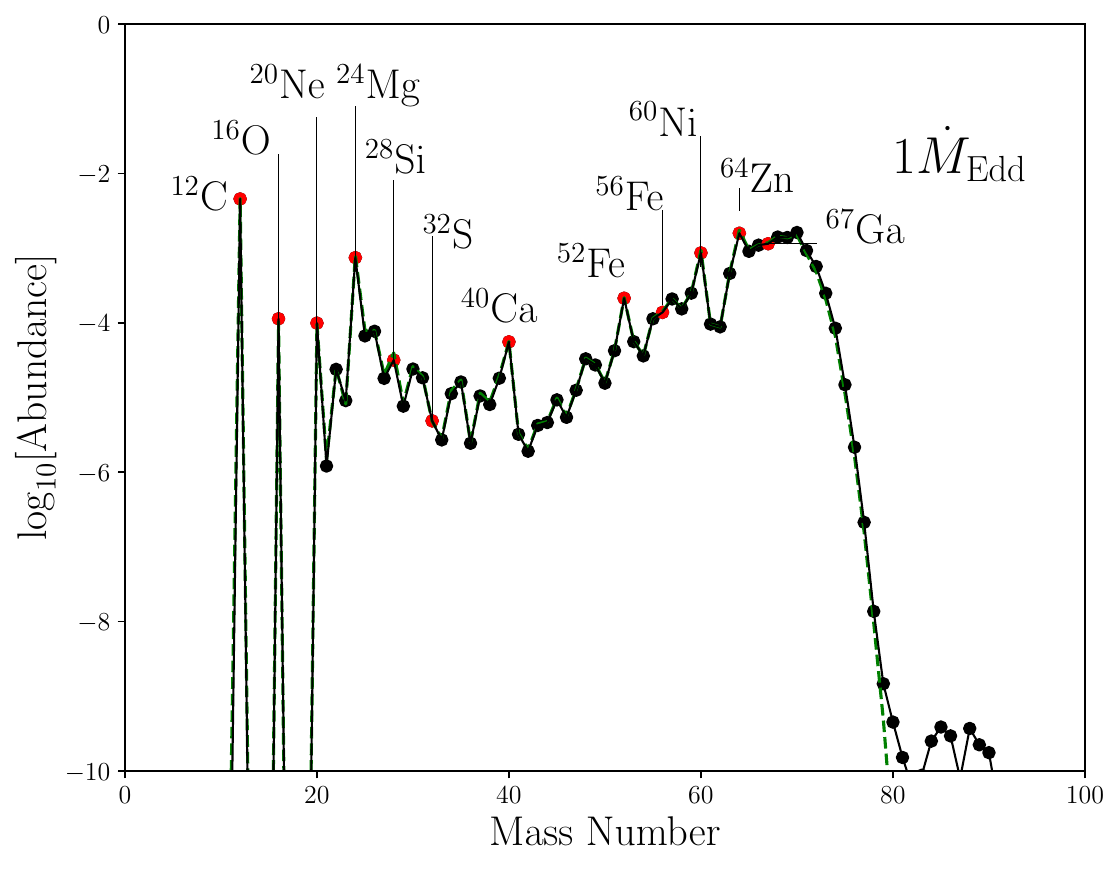}
 \caption{Distribution of abundances at $10^{7}$ g cm$^{-3}$ for $\dot{M}^{\infty} = \dot{M}_{\text{Edd}}$ as in the left panel of Figure~\ref{fig:abun_1_5mdot} (continuous black line)
 and a similar model in which the abundances of all accreted nuclei above He are set to zero (green dashed line).
 For some selected isobars, red dots, we indicate the most abundant nucleus.}
\label{fig:abun_1_noCNO}
\end{figure}

The above discussion shows that production of heavy elements is predominantly done at the burning of helium, which generates the $^{12}$C via the $3\alpha$-process.
To confirm this interpretation we compare in Figure ~\ref{fig:abun_1_noCNO} the resulting abundances of our $\dot{M}_{\text{Edd}}$ envelope with a similar model,
i.e. same $\dot{M}^{\infty}$, $T_\mathrm{eff}$, $\mtot$, and $\rtot$, but only accreting $^{1}$H and $^{4}$He with mass fractions 0.7 and 0.3, respectively.
One clearly sees that the final abundances are essentially identical, differing around and above $A=80$ only as a consequence of the lack of metals in the accreted composition. However, the similarity among $A<80$ ashes confirms all accreted CNO nuclei are consumed and the rp-process is fed almost entirely by the $^{12}$C produced by $^{4}$He burning.

In Table~\ref{tab:lums_eps} we report the luminosity at infinity from the processes illustrated in panels (d) and (d') of Fig.~\ref{fig:1_5medd_tx1}, computed with Eq.~\ref{eq:dif_lums2}.
As above, we infer the rp-process taking place between $A = 40$ and $A = 60$ represents the largest source of luminosity, in agreement with both peaks in Fig.~\ref{fig:1_5medd_tx1}, and the contributions from weak reactions near $\rho_{b}$. 
Considering a red-shift $e^{\Phi}\approx 0.8$ for these envelopes, at $\dot{M}_{\text{Edd}}$ we see the $(p,\gamma)$ and $(\beta^{+}\nu_{e})$ contributions from $A\geq 20$ amount to $\sim 2.27$ and $\sim 3.74$ MeV per nucleon respectively, i.e. the luminosity from nuclear reactions is $\approx$ 6 MeV per nucleon, similar to what is usually obtained via CNO burning at lower accretion rates. 
The scenario is slightly different at $5\dot{M}_{\text{Edd}}$: for the same individual contributions we now have $\sim 2.21$ and $\sim 4.02$ respectively, i.e. weak decays are slightly more energetic since we have more metal abundances at $5\dot{M}_{\text{Edd}}$ than at $\dot{M}_{\text{Edd}}$ (e.g. Fig.~\ref{fig:abun_1_5mdot}). 
However, the net released energy is still $\approx 6$ MeV per nucleon.

\begin{table}
	\centering
	\caption{Luminosities at infinity due to the processes in panels (d) and (d') of Fig.~\ref{fig:1_5medd_tx1}. These were computed employing Eq.~\ref{eq:dif_lums2}.}
	\begin{tabular}{|p{2.0cm}|p{1.8cm}|p{2.0cm}|}
        \hline
        Process & \multicolumn{2}{|c|}{$L^{\infty}$ [erg s$^{-1}$] for} \\
		 & $\dot{M}^\infty = \dot{M}_{\text{Edd}}$ & $\dot{M}^\infty = 5\dot{M}_{\text{Edd}}$\\
		\hline
		\hline
		Hot CNO & 1.57$\times 10^{35}$ & 3.79$\times 10^{35}$\\
		$3\alpha$ & 1.17 $\times 10^{35}$ & 4.37$\times 10^{35}$\\
		$(p,\gamma)$, $A\geq 20$ & 1.92$\times 10^{36}$ & 9.37$\times 10^{36}$\\
		$(p,\gamma)$, $A\geq 40$ & 1.11$\times 10^{36}$ & 5.55$\times 10^{36}$\\
		$(p,\gamma)$, $A\geq 60$ & 2.63$\times 10^{35}$ & 2.22$\times 10^{36}$\\
		$(\beta^{+}\nu_{e})$, $A\geq 20$ & 3.17$\times 10^{36}$ & 1.71$\times 10^{37}$\\
		$(\beta^{+}\nu_{e})$, $A\geq 40$ & 1.77$\times 10^{36}$ & 1.13$\times 10^{37}$\\
		$(\beta^{+}\nu_{e})$, $A\geq 60$ & 4.75$\times 10^{35}$ & 4.11$\times 10^{36}$\\
		\hline
	\end{tabular}\label{tab:lums_eps}
\end{table}

\subsection{Variations on the accreted amount of H}\label{subsec:variablexy}

The comparison of models at $\dot{M}_{\text{Edd}}$ carried out in Fig.~\ref{fig:abun_1_noCNO} confirms that hydrogen and helium abundances are the critical ingredients in the synthesis of $A>40$ metals via the rp-process.
In order to explore the actual impact of the mass fractions of $^{1}$H and $^{4}$He in the accreted matter composition, we performed a series of simulations varying their individual mass fractions, $X_{\text{H}}$ and $X_{^{4}\text{He}}$ respectively, while retaining their sum $X_{\text{H}} + X_{^{4}\text{He}}$ constant and equal to its value at Solar composition, $\approx$ 0.98. 
For all models, we employed the same $\mtot$, $\rtot$ as in Section \ref{subsec:res_medd} but, instead of fixing $T_{\text{eff}}$, we required all our models to have the same $T_{b}$ in order to have a fair comparison on the thermonuclear reaction rates. 

\begin{figure*}
\includegraphics[width=0.49\linewidth]{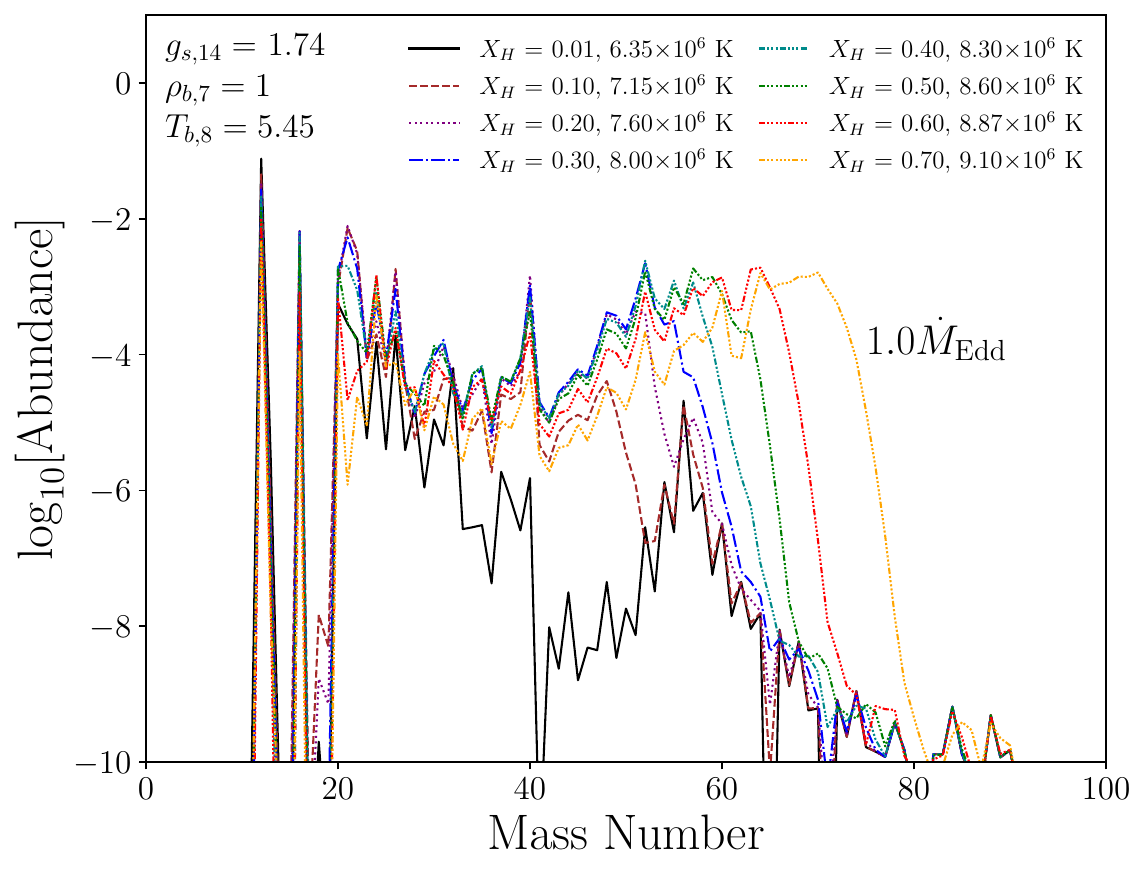}
\hspace{\stretch{1}}
\includegraphics[width=0.49\linewidth]{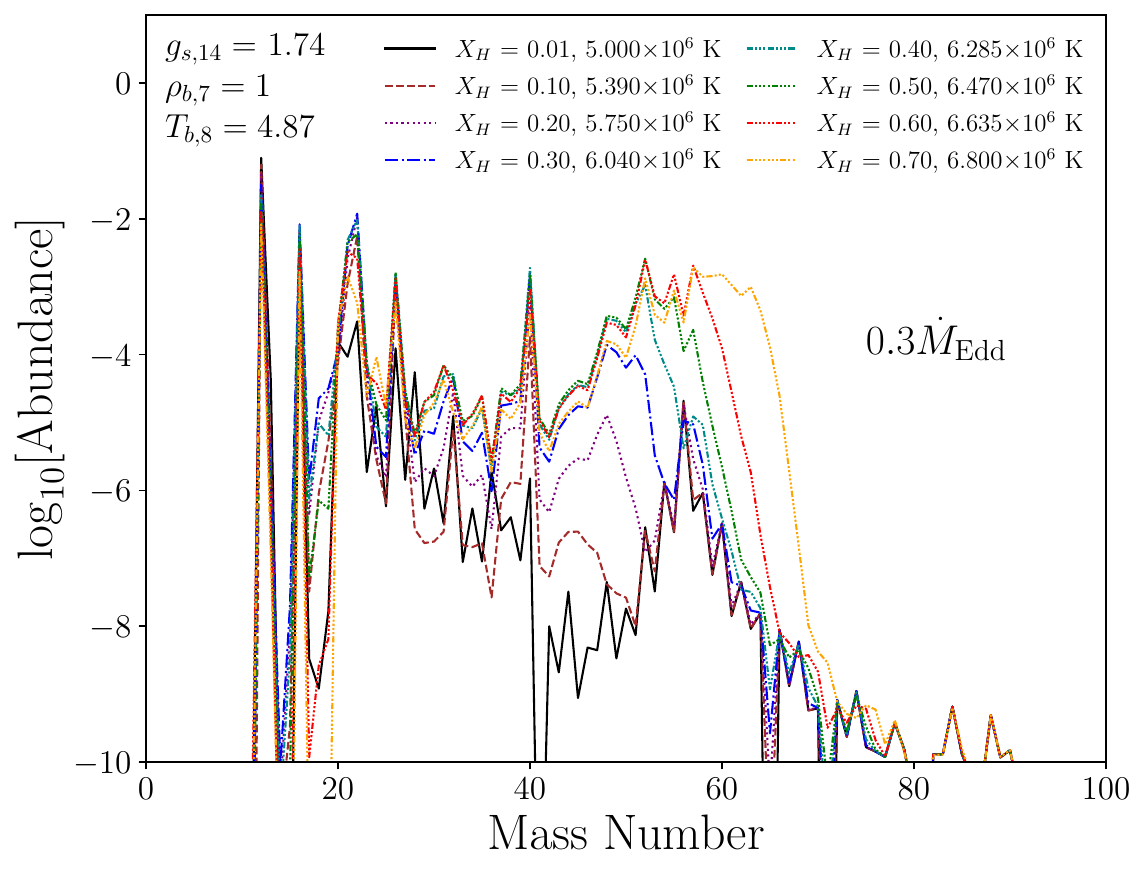}
\caption{Distribution of abundances for different accreted fractions of hydrogen, denoted as $X_\mathrm{H}$, and helium at the surface with $X_\mathrm{He} = 0.98-X_\mathrm{H}$. Left panel: $\dot{M}^{\infty} = \dot{M}_{\text{Edd}}$, right panel: $\dot{M}^{\infty} = 0.30\dot{M}_{\text{Edd}}$. Here, $\rho_{b,7} = \rho_{b}/10^{7}$ g cm$^{-3}$ and $T_{b,8} = T_{b}/10^{8}$ K.}
\label{fig:sev_lb_plot}
\end{figure*}

The resulting abundances for $\dot{M}_{\text{Edd}}$ at $\rho_{b}$ can be found in the left panel of Fig.~\ref{fig:sev_lb_plot}.
In the $X_{H} = 0.01$ scenario, nuclei at $A>40$ are not produced at all, as shown by their distribution, with a peak around $A = 56$ (i.e. $X_{^{56}\text{Fe}}\sim 1.2\times 10^{-3}$),
which is identical to the accreted matter composition.
Thus, the $A>40$ abundances for the $X_{H} = 0.01$ scenario is a reflection of the Solar metallicity. 
Between $X_{H} = $ 0.20 and 0.30 we see an ongoing rp-process as the abundance of $A = 52$ isobars becomes comparable and/or exceeds the abundance of $A = 40$. 
For 0.30 $\leq X_{^{1}\text{H}}\leq$ 0.70, we see the progressive increment of the fraction of accreted hydrogen allows to reach higher abundances for $A\geq 56$ isotopes. 
As clearly shown in these curves, the accreted hydrogen must exceeds 50\% of the total mass fraction in order to synthesize $A\geq 60$ material.


At 0.3$\dot{M}_{\text{Edd}}$, right panel of Fig.~\ref{fig:sev_lb_plot}, we find some similarities with the $\dot{M}_{\text{Edd}}$ case: $^{1}$H burning does not result in enhancement of $A> 40$ nuclides unless its mass fraction at the surface reaches $\sim 0.30$. 
However, we also observe that as long as $X_{^{1}\text{H}}$ constitutes up to $50\%$ percent of the accreted material, at this relative small accretion rate an rp-process takes place, synthesizing $A\sim 60$ nuclides
(since the $A>40$ composition at 0.01$\dot{M}_{\text{Edd}}$ is the original composition of the accreted matter, any excess above these values exhibits nuclei produced by the rp-process).

\subsection{rp-process at low-accretion rate}\label{subsec:lowmdotrp}

\begin{figure}
\includegraphics[width=\linewidth]
    {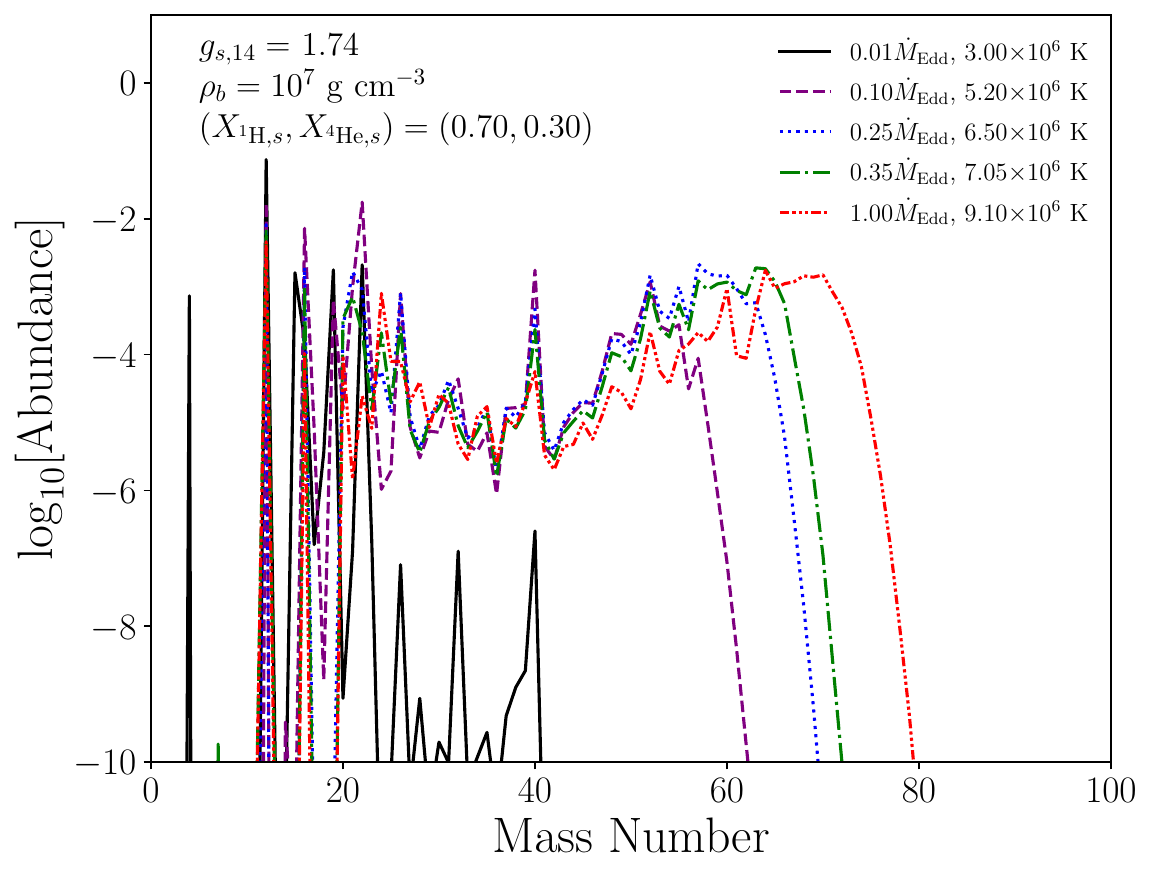}
 \caption{Distribution of abundances for different accretion rates, in units of $\dot{M}_{\text{Edd}}$.}
 \label{fig:sev_lowmdot}
\end{figure}

In this section we explore the efficiency of the rp-process as a function of $\dot{M}$ at low accretion rates.
For these models, we employed the same surface gravity as in Section~\ref{subsec:res_medd} and a simplified mixture of 0.70\% $^{1}$H and 30\% $^{4}$He, i.e. no metals at all since we have shown these have no significant impact.
The results are displayed in Fig.~\ref{fig:sev_lowmdot}, and for completeness in the discussion we also include the previously addressed $\dot{M}_{\text{Edd}}$ case.

At 0.01$\dot{M}_{\text{Edd}}$ we observe the disappearance of $^{1}$H that has been mostly converted into $^{4}$He by the CNO cycle, 
and a strong abundance of $^{12}$C as a results of the $3\alpha$ reaction.
Although synthesis beyond $A = 20$ is mostly negligible, we see some $A = 40$ material, as expected due to the bottleneck role of $^{40}$Ca in the rp-process.
At ten times this rate, 0.1$\dot{M}_{\text{Edd}}$, we observe an enhancement of this material reaching abundances up to $10^{-3}$, about an order of magnitude below the $A \sim 20$ material. 
However, the bottleneck of $^{40}$Ca is starting to be bypassed as we observe the synthesis of $A = 52$ and $A=56$ nuclei, with abundances an order of magnitude below that of $A=40$. 
Between 0.1 and 0.25$\dot{M}_{\text{Edd}}$ we finally see the complete bypass of this bottleneck since material up to $A = 60$ is synthesized. 
As we move from 0.25 to 0.35 to 1 times $\dot{M}_{\text{Edd}}$, the families of isobars $A = 60$ to $70$ are progressively synthesized since the temperatures for these stationary envelopes are progressively higher.

\begin{figure*}
\includegraphics[width=0.99\linewidth]{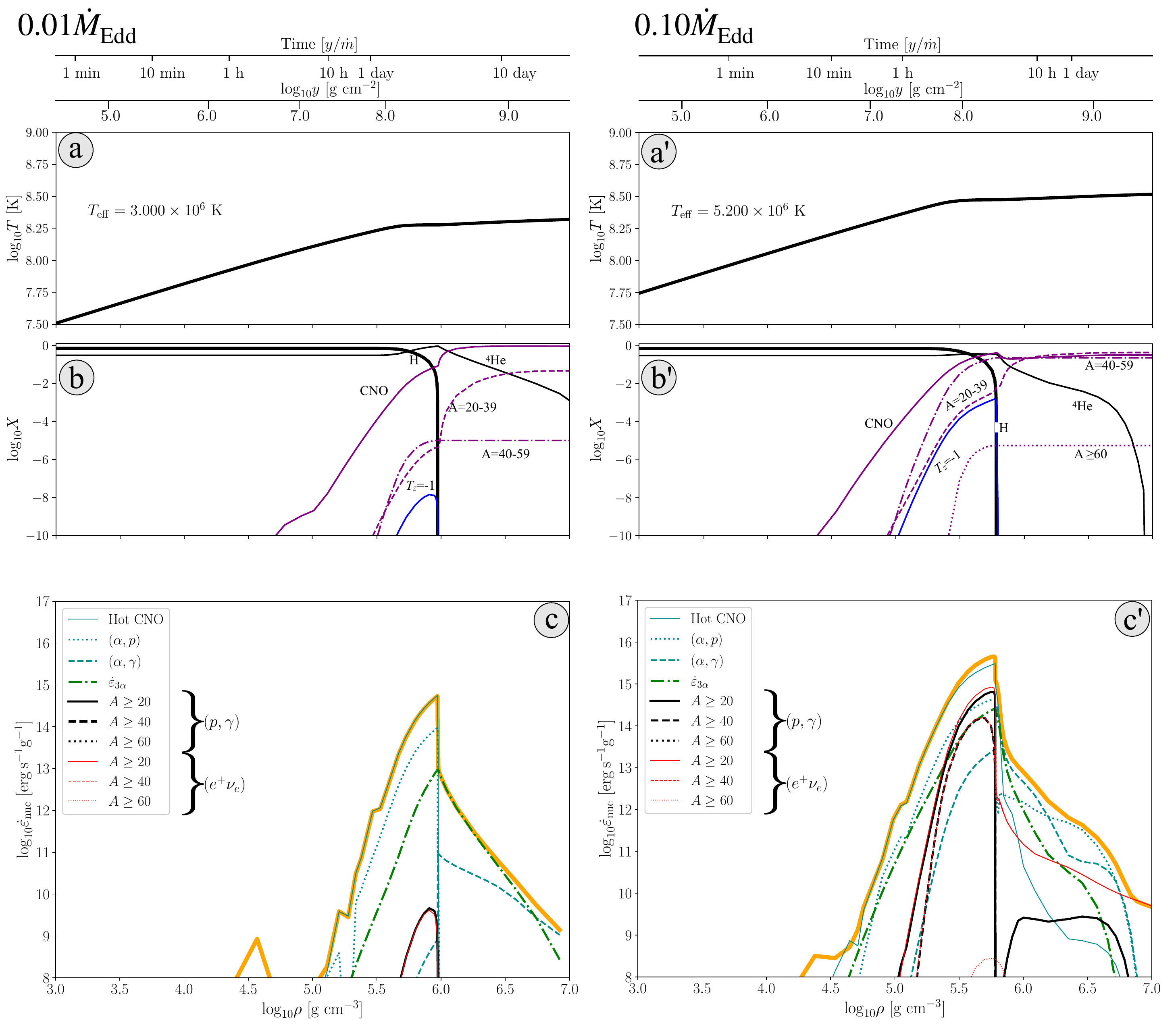}
\caption{Envelope models at $\dot{M}^{\infty} = 0.01\dot{M}_{\text{Edd}}$ (left panels) and $0.10\dot{M}_{\text{Edd}}$ (right panels) as a function of $\rho$. The upper scales indicate the corresponding column depth $y$ and time spent by the accreted matter since it started its journey from the neutron star surface.
Panels (a) and (a'): temperature;
panels (b) and (b'): mass fraction of selected nuclei; 
panels (c) and (c'): specific energy generation of dominant processes, as indicated, and with the upper thick line showing the total energy generation.}
 \label{fig:lowacsev1}
\end{figure*}

\begin{figure*}
\includegraphics[width=0.99\linewidth]{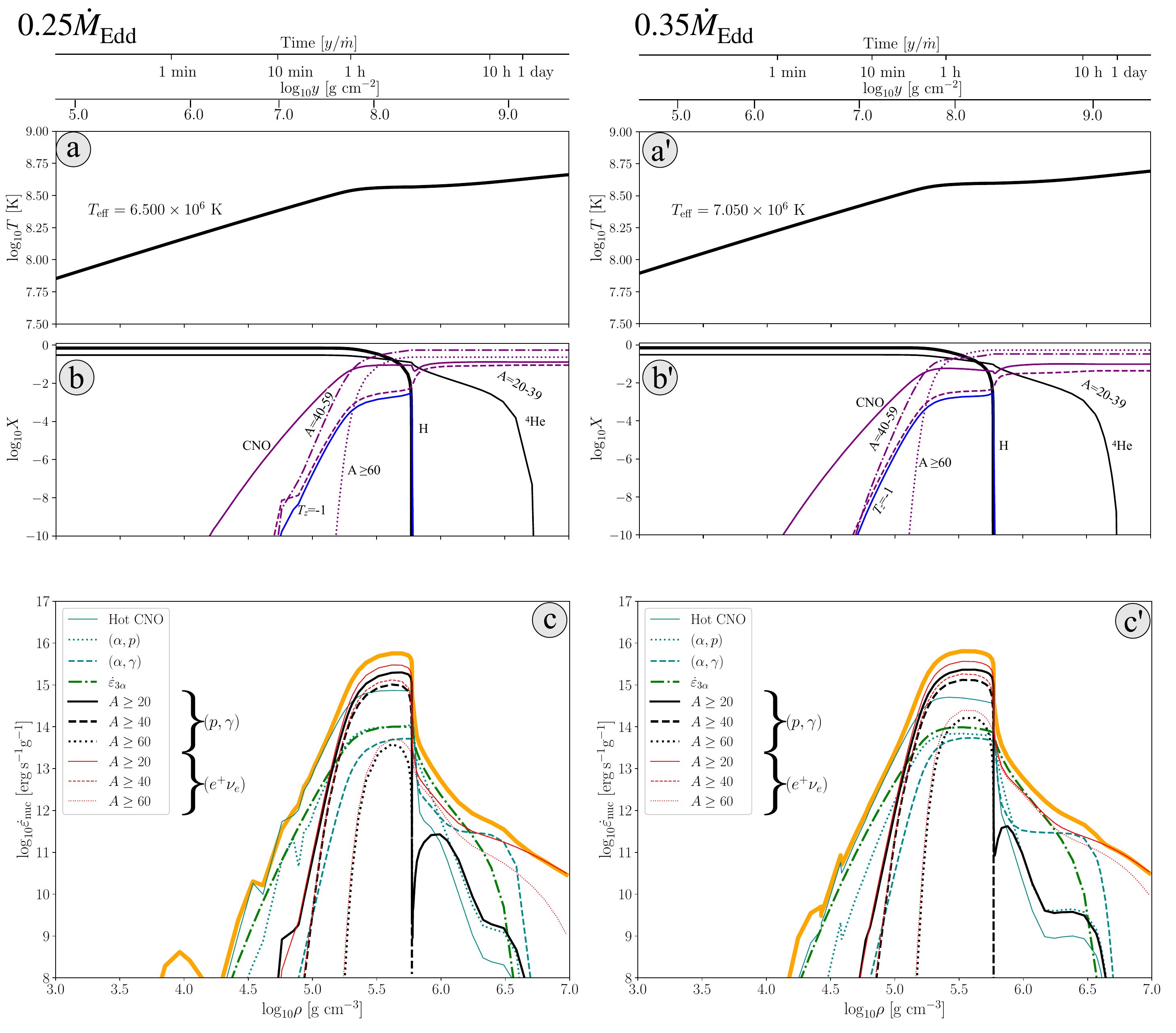}
\caption{Same as Fig. \ref{fig:lowacsev1}, for $\dot{M}^{\infty} = 0.25\dot{M}_{\text{Edd}}$ (left panels) and $0.35\dot{M}_{\text{Edd}}$ (right panels).}
 \label{fig:lowacsev2}
\end{figure*}

In Figs.~\ref{fig:lowacsev1} and~\ref{fig:lowacsev2}, we plot the corresponding envelope profiles. 
Due to the absence of metals in the accreted material, in all four cases the actual synthesis of nuclides occurs around and above $\text{log}_{10}\rho\sim 5$ and $\text{log}_{10}T \sim 8$, as conditions are suitable for 
the initiation of the $3\alpha$ reaction, and simultaneously the triggering of the hot CNO by the produced $^{12}$C.
The subsequent burning, however, naturally depends on the accretion rate. 
For 0.01$\dot{M}_{\text{Edd}}$, left panels in Fig.~\ref{fig:lowacsev1}, we have the standard scenario of CNO cycles followed by $(\alpha,p)$ reactions (including the $(p,\alpha)$ contributions such as $^{15}$N($p,\alpha$)$^{12}$C) and $3\alpha$ burning. 
Once hydrogen is exhausted, production of $^{4}$He ceases and the system is hot enough as to enhance the He-burning via $3\alpha$, the predominant reaction as indicated in panel (c), resulting in $^{12}$C being the most abundant species in the mixture. 
The enhancement of $20\leq A< 39$ material in this scenario is due the $3\alpha$ and $(\alpha,\gamma)$ reactions.
This is also shown by the small amount of $T_{z} = -1$ nuclides synthesized before the H-exhaustion point, occurring just below $\text{log}_{10}\rho\sim 6.25$, and the absence of strong $\beta^{+}$ decay contributions at high depths. 

At ten times this rate, 0.1$\dot{M}_{\text{Edd}}$ in the right-hand side panels of Fig.~\ref{fig:lowacsev1}, we see a moderate rp-process undergoing before the H-exhaustion point, now located near $\text{log}_{10}\rho\sim 6$.
Energetics is dominated by the hot CNO but there is some contribution from the rp-process on nuclei $A \geq 20$ with abundant production of nuclei in the $A\geq 40$ range seen in panel (b').
After H exhaustion the largest contribution to energy comes from $^{4}$He burning with $(\alpha,\gamma)$, and eventually $(\alpha, p)$, reactions dominating over the $3\alpha$.
As a result, the population of $A = 20-39$ nuclei significantly grows.
When density approaches $10^7$ g cm\mmm, only $\beta$-decay reactions remain.

Finally, let us address the stationary envelopes with rates around 0.3$\dot{M}_{\text{Edd}}$, i.e. 0.25 and 0.35 $\dot{M}_{\text{Edd}}$ in Fig.~\ref{fig:lowacsev2}. 
In contrast to the 0.10$\dot{M}_{\text{Edd}}$ model, the rp-process becomes very efficient with production of a large amount of $40\leq A < 60$ nuclides occurring at $\text{log}_{10}\rho$ between $\sim 5$ and 5.25,
and then up to $A\geq 60$ material at higher densities.
Regarding the generation of energy, we see the $(p,\gamma)$ reactions and their subsequent $\beta^{+}$ decays now liberate more energy than the $3\alpha$-process, by between one to two orders of magnitude, around the global maximum in both panels (c) and (c'). 
As a consequence of this burning, we see $^{1}$H exhaustion takes place at a similar density among the 0.1, 0.25 and 0.35$\dot{M}_{\text{Edd}}$ models, i.e. $\text{log}_{10}\rho \sim $6. 
After H exhaustion the largest contribution to energy comes from $^{4}$He burning from the $3\alpha$ with a significant contribution from the $(\alpha, p)$, competing with the beta-decays.
As a result, the population of $A = 20-39$ nuclei significantly grows.
When density approaches $10^7$ g cm\mmm, only $\beta$-decay reactions remain.

We must emphasize that, at the accretion rates considered in this section,
our results at densities above $\sim 10^6$ g cm\mmm\ are likely purely academic. 
Indeed, since burning will become unstable  when the accreted matter reaches densities $\sim 10^6$ g cm\mmm,
the composition past this point should be described by X-ray burst ashes. On the other hand,
our description of matter at lower densities corresponds to the one in-between bursts.

\subsection{Helium accreting envelopes}\label{subsec:he_env}

\begin{figure*}
\includegraphics[width=0.45\linewidth]{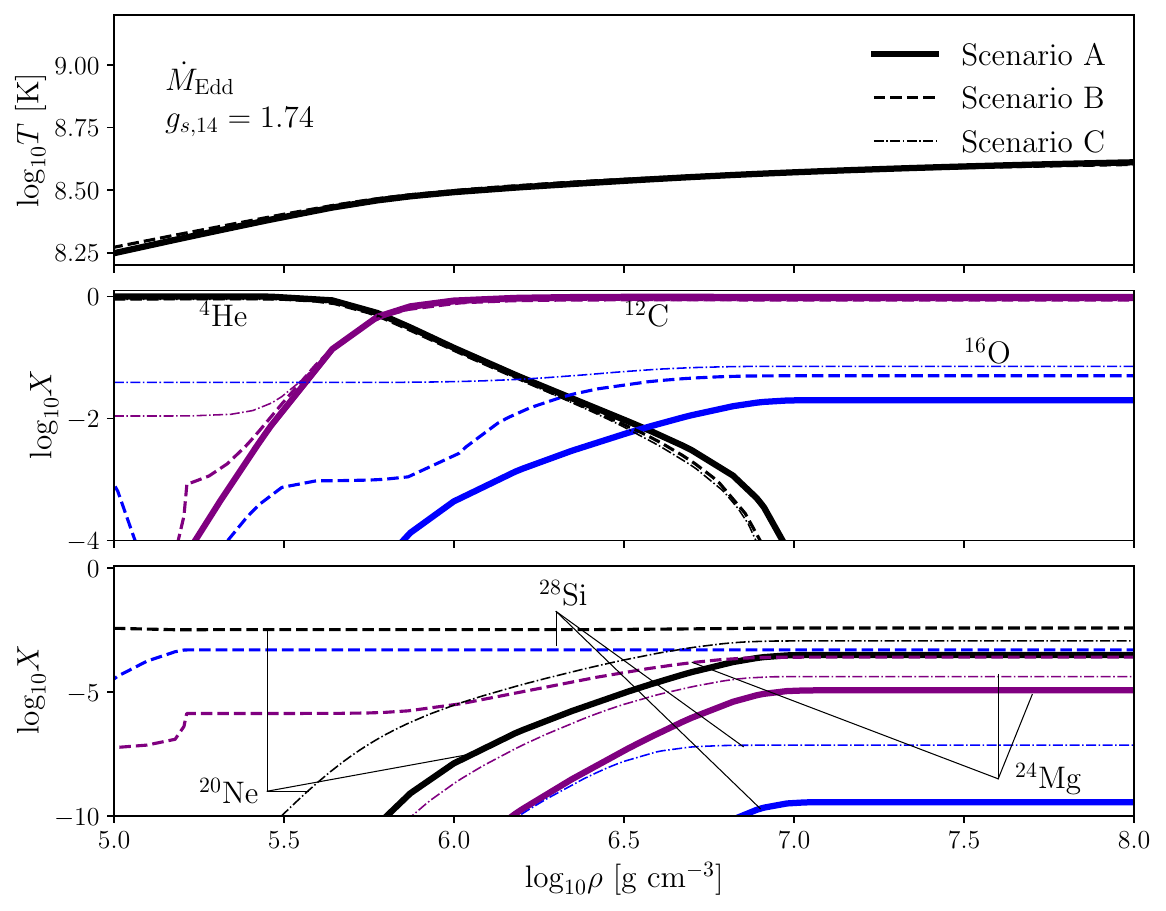}
\includegraphics[width=0.45\linewidth]{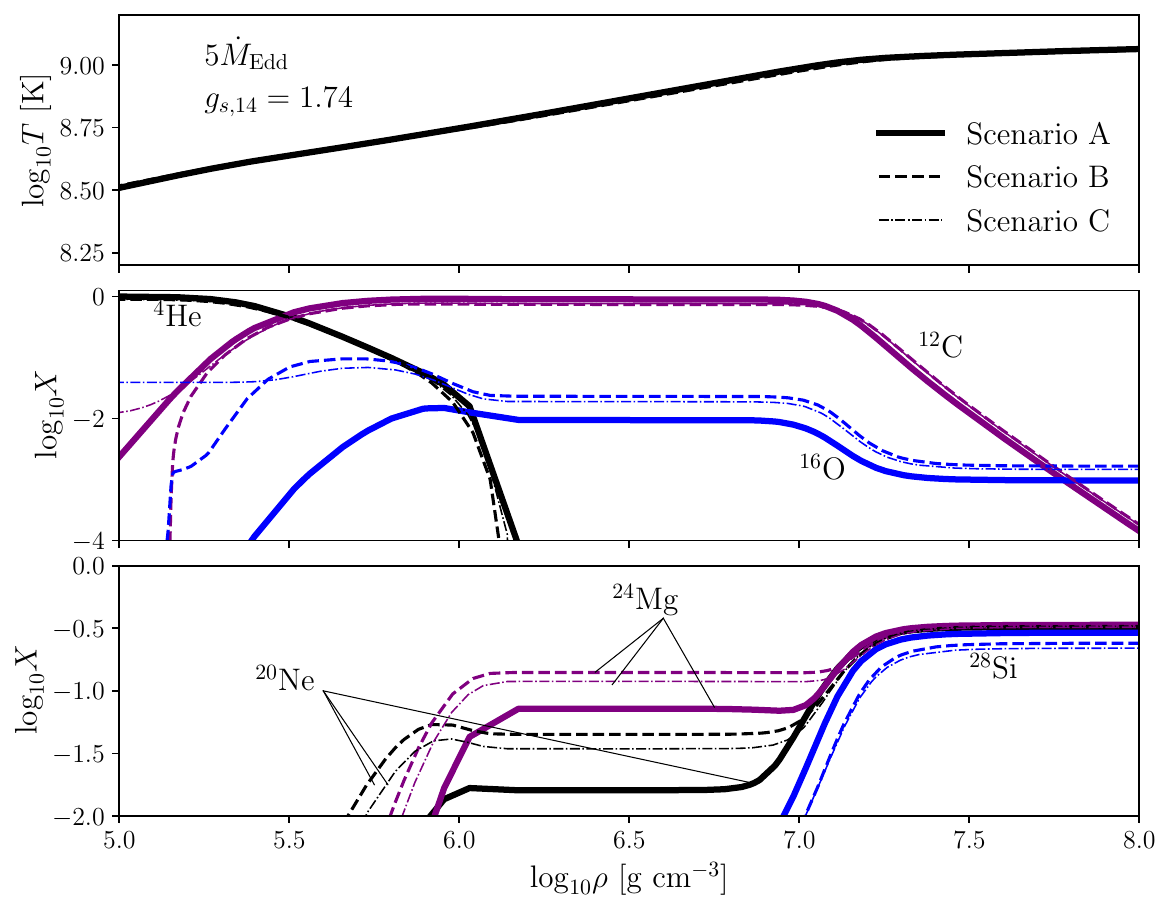}
 \caption{$^4$He accreting envelopes at $\dot{M} = \dot{M}_{\text{Edd}}$ (left panels) and $\dot{M} = 5\dot{M}_{\text{Edd}}$ (right panels) as a function of $\rho$. 
 Top panels: temperature. Low and middle panels: mass fractions of synthesized $\alpha$-nuclides. 
 Thickest to thinnest lines correspond to scenarios A, B and C respectively.}
 \label{fig:mostlyhe4_1medd}
\end{figure*}

\begin{figure*}
\includegraphics[width=0.45\linewidth]{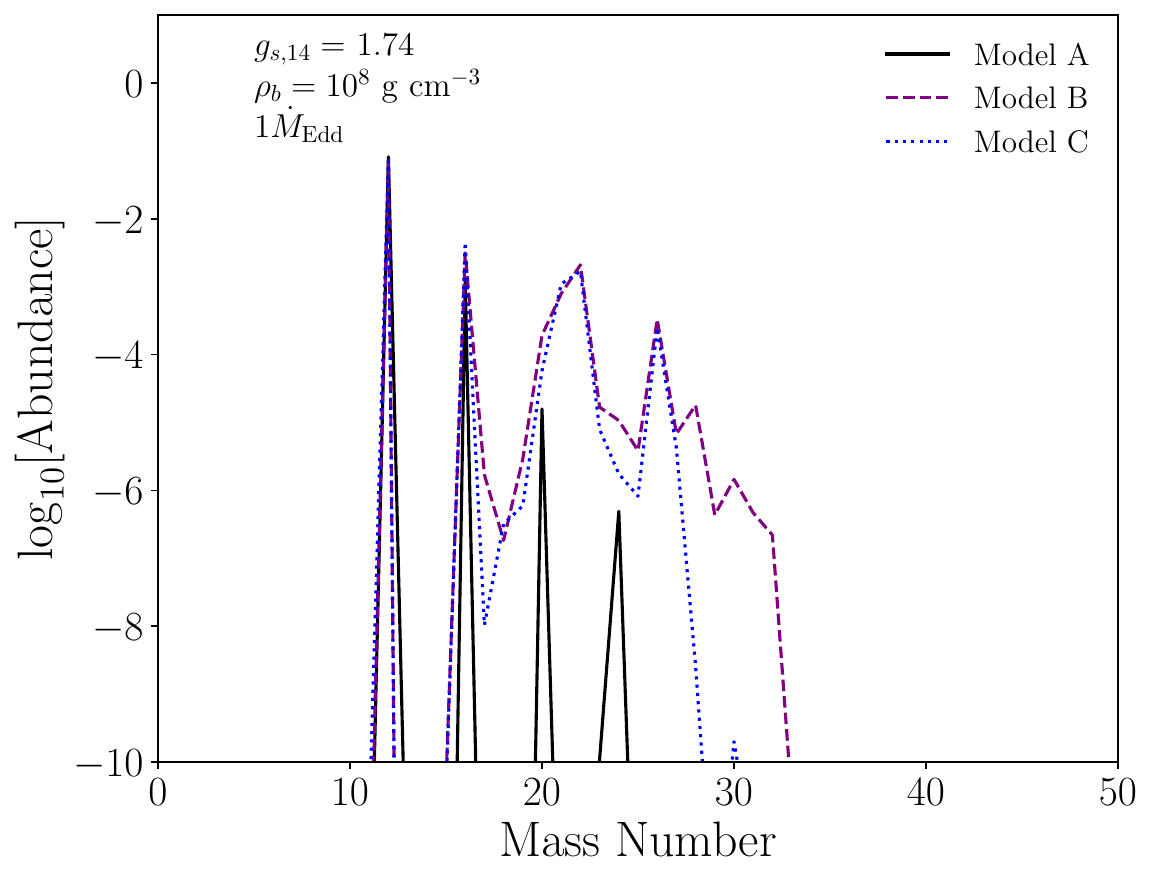}
\includegraphics[width=0.45\linewidth]{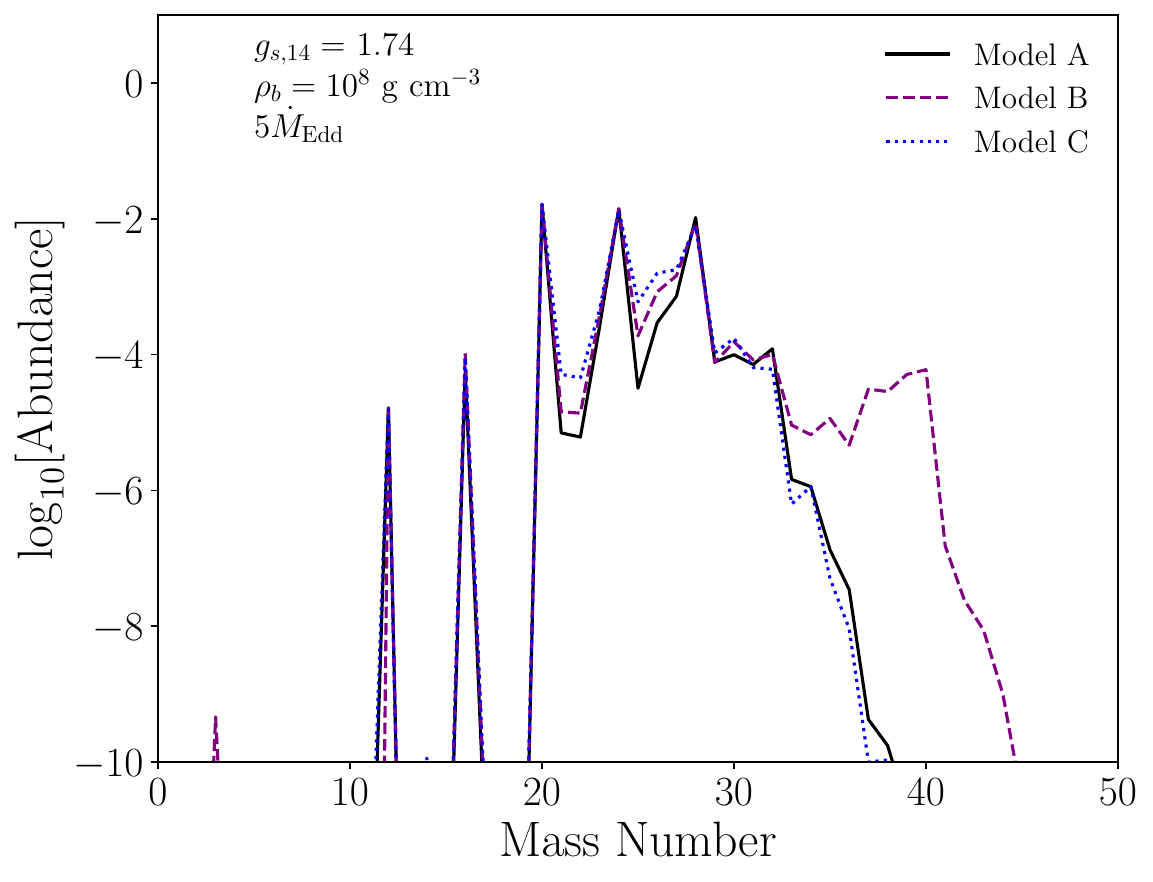}
 \caption{Distribution of abundances at $\rho_{b}$ for the $^4$He accreting envelopes at $\dot{M} = \dot{M}_{\text{Edd}}$ (left panels) and $\dot{M} = 5\dot{M}_{\text{Edd}}$ (right panels).}
 \label{fig:mostlyhe4_diaabu}
\end{figure*}

A completely different evolution is naturally expected when accreted matter is strongly hydrogen deficient.
We have tested three scenarios with large amounts of $^{4}$He at the surface, in order to determine how large can $\alpha$-nuclide abundances be as a consequence of accreting and burning $^{4}$He rich material. Taking $g_{s}$ as that for a $1.4 M_{\odot}$ star, employing two high accretion rates, 1 and 5 times $\dot{M}_{\text{Edd}}$, the proposed scenarios are:
\begin{itemize}
    \item \textbf{Scenario A:} $X_{^{4}\text{He}} = 1$.
    \item \textbf{Scenario B:} $X_{^{4}\text{He}} = 0.90$, $X_{^{12}\text{C}} = 0.01$,   $X_{^{14}\text{N}} = 0.051$, $X_{^{16}\text{O}} = 0.015$, with a small admixture of $X_{^{20}\text{Ne}} = 0.004$, and some hydrogen $X_{^{1}\text{H}} = 0.02$, 
    \item \textbf{Scenario C:} $X_{^{4}\text{He}} = 0.90$, $X_{^{12}\text{C}} = 0.011$, $X_{^{14}\text{N}} = 0.05$,   $X_{^{16}\text{O}} = 0.039$ (without H and Ne).
\end{itemize}
In the first scenario, we explore how much $\alpha$-nuclides can be synthesized by pure $^{4}$He burning, provided the high accretion rates guarantee sufficiently high temperature as to allow the operation of $3\alpha$ and $(\alpha,\gamma)$ reactions. In the second we keep a minimum amount of hydrogen and metals, sufficient to guarantee both $X_{^{4}\text{He}}/4 \gg X_{^{1}\text{H}}$ and the operation of the CNO cycles. 
The final scenario tests the influence of such CNO isotopes in the synthesis of $A\geq 20$ $\alpha$-nuclides provided abundant $^{4}$He accretion. 
This should also serve as a direct comparison with the first scenario. 
To guarantee a fair comparison among the scenarios, compensating for their different $T_{\text{eff}}$, we have fixed the base luminosity $L_b$ at $4\times 10^{34}$ erg s$^{-1}$ for the $\dot{M}_{\text{Edd}}$
models and at $2\times 10^{35}$ erg s$^{-1}$ in the $5 \dot{M}_{\text{Edd}}$ ones, i.e., assuming a flux from the interior proportional to the mas accretion rate.
The resulting $T$ and $\alpha$-nuclide mass fraction profiles can be seen in Fig.~\ref{fig:mostlyhe4_1medd} and the final abundances are presented in Fig.~\ref{fig:mostlyhe4_diaabu}.
For both accretion rates we observe temperature profiles following close trajectories in the $\rho$ vs $T$ plane. 
The same superposition is observed for $^{4}$He and $^{12}$C profiles above $10^{5.75}$ g cm$^{-3}$ at $\dot{M}_{\text{Edd}}$ and $10^{5.5}$ g cm$^{-3}$ at $5\dot{M}_{\text{Edd}}$, from where we infer the $3\alpha\to ^{12}$C rate governs this high-density sector of the envelope due to the large fraction of accreted helium. 

At $\dot{M}_{\text{Edd}}$, scenario A mass fractions for the rest of the species are the smallest among the three scenarios. 
The opposite situation occurs in scenario B, where the tiny amount of $^{1}$H  enhances the synthesis of $^{24}$Mg and $^{28}$Si, but no impact is observed over $^{20}$Ne. 
If hydrogen is absent, but we accrete some metals, scenario C, we still synthesize significant amounts of $^{20}$Ne, $^{24}$Mg and even some $^{28}$Si. 
Despite these differences at $\dot{M}_{\text{Edd}}$, the three scenarios agree on that $^{12}$C, followed by $^{16}$O, practically constitutes $\sim 90\%$ of the ashes, yielding similar abundances of $A=24$ and $A=28$ $\alpha$-nuclides as those obtained from accretion of Solar-like matter (e.g. Sec.~\ref{subsec:res_medd}). 

At $5\dot{M}_{\text{Edd}}$, due to higher temperatures, the nuclear burning is very different.
At densities around $10^6$ g cm\mmm{} we see the disappearance of $^4$He due to $\alpha$-capture reactions that generate some $^{20}$Ne and $^{24}$Mg.
Later, at densities around $10^7$ g cm\mmm, $^{12}$C-$^{12}$C and $^{12}$C-$^{16}$O fusion reactions, absent at the lower mass accretion rate, result in the
production of large amount of $^{20}$Ne, $^{24}$Mg and $^{28}$Si.
For the three scenarios, despite the different composition of accreted matter, we find that $^{24}$Mg, $^{20}$Ne and $^{28}$Si comprise $\sim 90\%$ of the total composition of the ashes, with small variations in their distribution at $\rho_{b}$, indicating the synthesis and burning of $^{16}$O are the actual waiting point of the burning process. 


\section{Conclusion}

In this paper we have introduced and validated our code for steady state envelopes, which we plan to use coupled with a whole thermal evolution code to explore phenomena such as shallow heating and the chemical composition of the accreted neuron star crust.
Our conclusions can be thus summarized:

\begin{itemize}
    \item Our numerical code provides reliable results, in comparison with those from the more robust network of $\sim$ 600 nuclides of \citet{1999ApJ...524.1014S}.
    The discrepancies, even considering the updated versions of some rates, are actually small. 
    The absence of $\rho, T$-dependent weak rates is more evident at high accretion rates due to the lesser amounts of synthesized $A=80$ - $90$ nuclides.
    \item We have explicitly shown how energy is distributed in these stationary envelopes.
     The rp-process can be distinguished by a prominent peak, the sum of $(p,\gamma)$ and $\beta^{+}$ decay energies just before H depletion, around $10^{6}$ g cm$^{-3}$. 
     Past this threshold, the energy mostly comes from $\beta$-decays and the $3\alpha$ reaction. 
    \item We saw a clear change in H burning occurring between  $\dot{M} = 0.1 \dot{M}_{\text{Edd}}$ and $0.2 \dot{M}_{\text{Edd}}$.
    Below $0.1 \dot{M}_{\text{Edd}}$ H burns almost completely through the hot CNO chain (Fig.~\ref{fig:lowacsev1}) while at $0.25 \dot{M}_{\text{Edd}}$ and $0.35 \dot{M}_{\text{Edd}}$ its
    burning is strongly dominated by rp-proccesses (Fig.~\ref{fig:lowacsev2}).
    As a byproduct we see a clear effect of the $^{40}$Ca bottleneck: at low accretion rates, as in our  $0.01\dot{M}_{\text{Edd}}$ model, essentially no nuclei above $A=40$ are synthesized, while at higher rates the rp-process is progressively generating increasing amounts of increasingly heavier nuclei as in our $0.1$, $0.25$ and $0.35\dot{M}_{\text{Edd}}$ models.
    \item Our envelope models being stationary, they only apply when nuclear burning is stable. 
    No X-ray burst have been observed at high accretion rates implying that our $\dot{M} = \dot{M}_{\text{Edd}}$ models are likely realistic and describe the structure and composition of the envelope
    in their whole density range.
    In the presence of explosive burning, at lower accretion rates, our models should give a good representation of the envelope only at densities below $\sim 10^6$ g cm\mmm, while the higher density regions should actually consist of X-ray burst ashes.
    \item We did not discuss the stability of our models considering it is better assessed with time dependent simulations and will be part of future works.
    \item  The $^{4}$He accretion at $5\dot{M}_{\text{Edd}}$ mainly seems of academic interest since, so far, observed He-accretion systems operate well below the Eddington rate. Nevertheless, our calculations suggest it is an interesting scenario as the amounts of $A=24$, $A=28$ material left at deeper layers might favor the occurrence of hyperbursts \citep{Page_2022}, a thermonuclear explosion requiring $\alpha$-nuclides of $A=24$ or $28$. \citet{Page_2022} inferred that hyperbursts are very rare events as they are triggered at around $10^{11}$ g cm\mmm, while our results in the cases of $^4$He accretion at high rate suggest that under the right conditions the waiting time may be shorter.
 \end{itemize}

\section*{Acknowledgements} 

M. Nava-Callejas thanks A. Cumming for discussions on the code and references, and F. Timmes for the pedagogic resources on the construction of nuclear networks at \url{https://cococubed.com/code_pages/burn.shtml}. YC acknowledges support from the grant RYC2021-032718-I, financed by MCIN/AEI/10.13039/501100011033 and the European Union NextGenerationEU/PRTR.

\bibliographystyle{mnras}
\bibliography{References} 




\appendix

\section{On the envelope equations}\label{sec:odes_extended}

\textbf{Structure.} The spacetime of very slow- or non-rotating stars can be modeled as static and spherically symmetric, according to the line element in spherical coordinates $(ct,r,\theta,\phi)$
\begin{equation}
ds^{2} = -e^{2\Phi(r)}c^{2}dt^{2} + e^{2\Lambda(r)}dr^{2} + r^{2}\left(d\theta^{2} + \sin^{2}\theta d\phi^{2}\right)
\end{equation}
with $e^{\Phi(r)}$ being the \textit{redshift}, and $\Lambda(r)$ being related to the gravitational mass via $m(r) = c^{2}r(1 - e^{-2\Lambda(r)})/2G$. By inserting this metric into Einstein field equations and using the constraint for the energy-momentum tensor $\nabla_{\mu} T^{\mu\nu} = 0$, one obtains \citep{Shapiro:1983wz}
\begin{align}
\frac{dm}{dr} & = 4\pi \rho r^{2}\\
\frac{dP}{dr} & = -g\, e^{\Lambda}\mathcal{H}\mathcal{G}\\
\frac{d\Phi}{dr} & = -\frac{1}{c^{2}\mathcal{H}}\frac{dP}{dr},
\end{align}
where $\mathcal{G}$ and $\mathcal{H}$ are given by Eqs.~\ref{eq:thorne_1} and ~\ref{eq:thorne_2}.

\textbf{Thermal evolution.} This category is governed by three aspects: composition, temperature and energy transport. In the presence of radial mass accretion, the fluid 4-velocity $\vec{u}$ can be expressed as $\vec{u}$ = $\gamma_{S}$ ($e^{-\Phi}$, $ve^{-\Lambda}$, $0$, $0$), in terms of the 3-velocity $v<0$ and with $\gamma_{S} = (1 - v^{2}/c^{2})^{-1/2}$. Together with the continuity equation $\nabla_{\mu}(n_\mathrm{B}u^{\mu}) = 0$, with $n_\mathrm{B}$ and $\rhobaryon = m_{p}n_\mathrm{B}$ the baryon number and rest-mass densities respectively, $m_p$ being the proton mass, in the stationary case we have a conserved quantity: $\dot{M}^{\infty} := -4\pi r^{2}e^{\Phi}\rhobaryon\gamma_{S}v$. Near the surface of the star, at $\dot{M}_{\text{Edd}}$, we see $\gamma_{S}|v|/c \sim 10^{-3}$, i.e. $|v|\sim \times 10^{-3}$c. Due to $\gamma_{S}|v|/c\propto\rho^{-1}$, $|v|\ll c$ at high depths. Therefore, $\gamma_{S}\approx 1$ in the whole envelope. 
Above the neutron star surface where accreting matter is in almost free-fall, $\rhobaryon$ is much smaller and $v$ close to $c$.

In general, we can have a mixture of nuclides. Each species is characterized by: (i) two integers, charge and neutron numbers $Z_{i}$, $N_{i}$, (ii) its binding energy $\mathcal{B}_{i}$, (iii) its abundance $Y_{i} = n_{i}/n_{B}$ (with $n_{i}$ being its number density) and (iv) their corresponding partial differential equation
\begin{equation}
u^{\mu}\partial_{\mu}Y_{i} = \mathcal{R}_{i}(T,\rho,\bm{Y}) \; , 
\label{eq:abundances}
\end{equation}
where the right-hand side functional encloses all i-th species related creation/annihilation rates. The whole set of equations $u^{\mu}\partial_{\mu}\bm{Y}$ is known as the \textit{nuclear reaction network} \citep{1999ApJS..124..241T, 2006NuPhA.777..188H}, and the released energy per gram is given by
\begin{equation}
\dot{\varepsilon}_{\text{nuc}} = -\sum_{j}N_{A}M_{j}c^{2}\mathcal{R}_{j} + \sum_{k\in\mathcal{W}}\langle \dot{e}_{\nu}\rangle Y_{k},
\label{eq:eps_nuc_ss}
\end{equation}
where $N_{A}$ is the Avogadro's constant, $M_{i}c^{2} = Z_{i}(m_{p} + m_{e})c^2 + N_{i}m_{n}c^2 - \mathcal{B}_{i}$, with $m_{n}$ and $m_{e}$ being the rest-masses of the neutron and the electron, respectively. $\mathcal{W}$ denotes the nuclide subset undergoing electroweak reactions, and $\langle \dot{e}_{\nu}\rangle$ denotes the average specific energy per unit time carried away by neutrinos\footnote{We 
reserve the notation $ \dot{\varepsilon}_{\nu}$ for the thermal neutrinos energy losses, see \S~\ref{sec:microphysics}}. 
To quantify the abundance fraction flowing through a certain channel \textit{directly} connecting nuclides $i$ and $j$, the integrated reaction flow $F_{ij}$ is introduced. 
In our time-independent approximation, the appropriate expression is
\begin{equation}
F_{ij} = \int^{P_{b}}_{P_{s}}dP\, \left[\frac{dY_{i}}{dP}(i\to j) - \frac{dY_{j}}{dP}(j\to i)\right],
\label{eq:Fij}
\end{equation}
where $\frac{dY_{A}}{dP}(A\to B)$ denotes the portion of species $A$ ordinary differential equation linking $A$ with $B$. 

The temperature evolution is implicitly given by 
\begin{equation}
\frac{\partial(e^{2\Phi}L)}{\partial r}= 4\pi r^{2}e^{2\Phi + \Lambda}\rhobaryon\left(\dot{\varepsilon}_{\text{nuc}} + \dot{\varepsilon}_{\text{grav}} - \dot{\varepsilon}_{\nu}\right),
\label{eq:dif_lums1}
\end{equation}
where $\dot{\varepsilon}_{\text{grav}} := -Tu^{\mu}\partial_{\mu}\breve{s}$, known in \texttt{MESA} as the gravitational energy, is best handled by making a distinction between the purely temporal (non-homologous) and spatial (homologous) contributions \citep{1975PASJ...27..197S, Paxton_2015}. In our stationary approximation, the first is automatically zero and the second can be written as \citep{1984ApJ...278..813F, Townsley_2004, Paxton_2015}
\begin{equation}
\dot{\varepsilon}_{\text{grav-h.}} = -\frac{\dot{M}^{\infty}e^{-\Phi - \Lambda}\breve{c}_{P}T}{4\pi r^{2}\rhobaryon P}\frac{dP}{dr}\left(\nabla_{\text{ad}} - \nabla_{T}\right),
\end{equation}
with $\nabla_{T} = \frac{P}{T}dT/dP$ and $\nabla_\mathrm{ad}$ the thermodynamic adiabatic gradient, i.e. the same expression but with $dT/dP$ taken under an adiabatic transformation.
Since $\nabla_{\text{ad}}>\nabla_{T}$, $\dot{\varepsilon}_{\text{grav, h.}} > 0$. 
%
%
%
The relation between the gravitational and nuclear rates significantly depend on $\rho$: below $\sim 10^{3}$ g cm$^{-3}$, where matter is an almost ideal gas ($\breve{c}_{P}\approx 5P/2\rho T$), a conservative estimation leads to
\begin{equation}
\dot{\breve{\varepsilon}}_{\text{grav-h.}}\approx  \left(\frac{10^{3}\, \text{g cm}^{-3}}{\rhobaryon}\right) \left(\frac{\dot{M}^{\infty}}{\dot{M}^{\infty}_{\text{Edd}}}\right)\times 10^{15.5}\ \text{erg s}^{-1}\text{g}^{-1},
\label{eq:egravenuc1}
\end{equation}
which far exceeds any contribution from nuclear reactions (e.g. $pp$ chains or initial phases of the CNO cycle). As $\rho\gg 10^{3}$ g cm\mmm, within the degenerate electron domain, $\dot{\breve{\varepsilon}}_{\text{grav-h}}\propto \rho^{-4/3}$ and thus $\dot{\varepsilon}_{\text{nuc}}$ $\gg$ $\dot{\varepsilon}_{\text{grav}}$.

Considering the nuclear energy is a sum of contributions from the released energy of the individual reaction rates (e.g. Eq.~\ref{eq:eps_nuc_ss}), we can compute their individual contribution to $L^{\infty}$ using Eq.~\ref{eq:dif_lums1} as
\begin{equation}
L^{\infty}_{j} = \int^{r_{\ast}}_{r_{b}}dr\, 4\pi r^{2}e^{2\Phi + \Lambda}\rhobaryon\dot{\varepsilon}_{j}.
\label{eq:dif_lums2}
\end{equation}
Finally, energy transport is treated via
\begin{equation}
L =  -4\pi r^{2}Ke^{-\Phi-\Lambda}\frac{\partial}{\partial r}(e^{\Phi}T),
\end{equation}
where the thermal conductivity, $K = \frac{16\sigmasb T^{3}}{3\rhobaryon\kappa_{\text{eff}}}$, includes both radiation and conduction effects in the effective opacity $\kappa_{\text{eff}} = [\kappa^{-1}_{\text{cond}} + \kappa^{-1}_{\text{rad}}]^{-1}$
with $\kappa_\text{cond}$ and $\kappa_\text{rad}$ defined below.

\begin{figure*}
\includegraphics[width=0.49\linewidth]{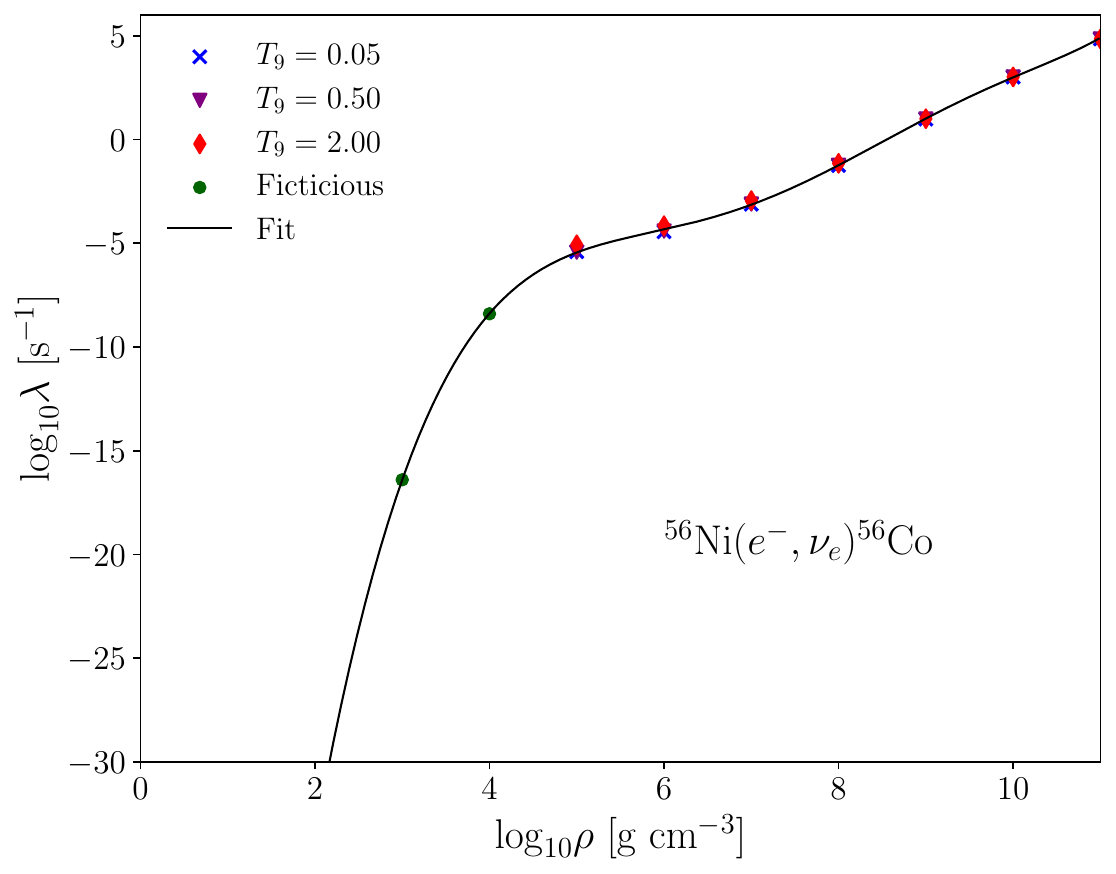}
\hspace{\stretch{1}}
\includegraphics[width=0.49\linewidth]{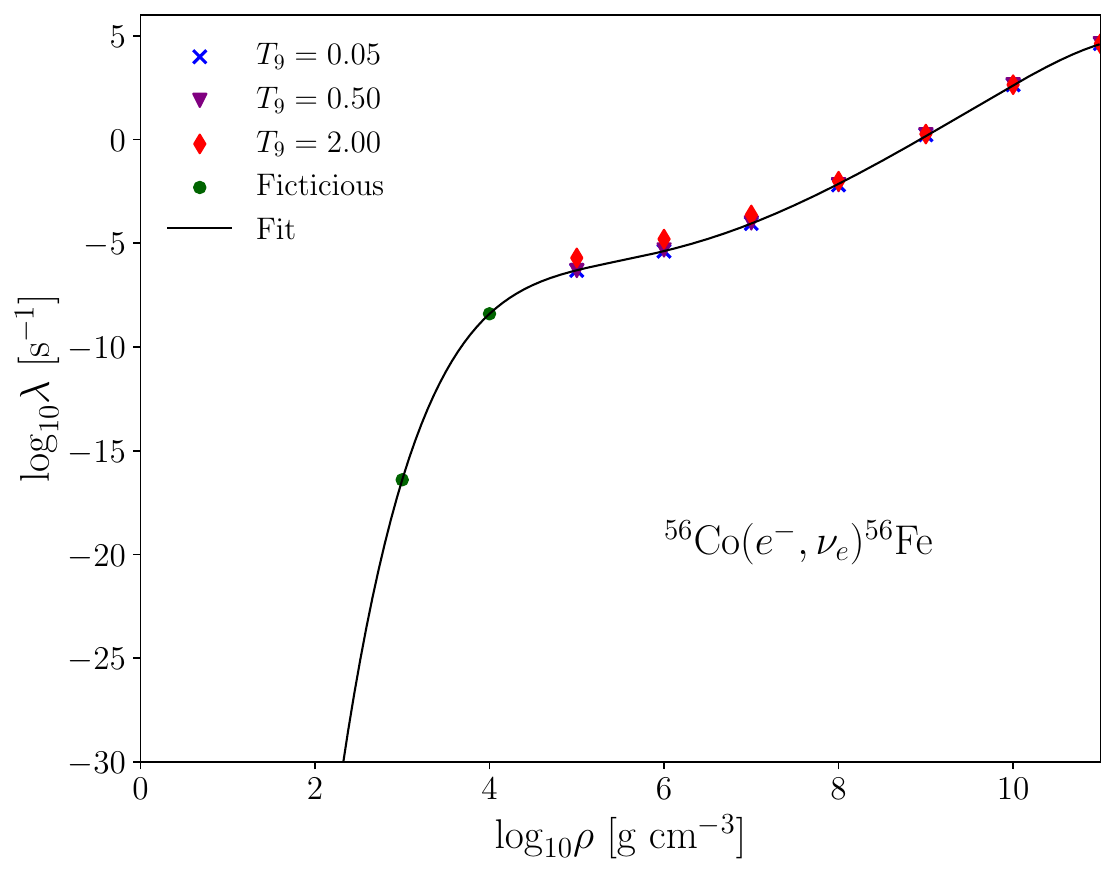}
\caption{Electron capture reaction rate for $A=56$ nuclides.}
\label{fig:ec_a56}
\end{figure*}

\section{Microphysical input}\label{sec:microphysics}

\textbf{The Equation of State (EOS).} We adopt the standard non-interacting EOS of stellar matter \citep{2012sse..book.....K}, subject to charge neutrality. For electron pressure we used the fits from \cite{1996ApJ...473.1020J}. To test the impact of electrostatic interactions we compared all the results from this EOS to those obtained by \citet{PhysRevE.62.8554, pc_eos_ref1} and  \citet{Haensel:2007un}, employing the fits from \cite{Haensel:2007un, ichimaru198791, PhysRevE.58.4941, PhysRevE.62.8554}. Since the observed differences are smaller than $10\%$ in temperature, in the present work we can safely neglect these corrections in favor of speeding up the numerical calculations.

\textbf{Opacity.} In the Rosseland mean $\kappa_{\text{rad}}$ we include the electron scattering expression from \cite{2017ApJ...835..119P}, free-free absorption as described in \cite{1999ApJ...524.1014S}, and the correction factor from \cite{2001A&A...374..213P}. For $\kappa_{\text{cond}}$, we consider the electron-electron scattering from \cite{2006PhRvD..74d3004S}, and the fit from \cite{1999ApJ...524.1014S} for electron-ion scattering.

\textbf{Thermal neutrinos.} We include them considering the fits of \cite{1996ApJS..102..411I} for 5 processes: plasmon decay, photo- and pair-neutrinos, recombination and bremsstrahlung. The latter contribution was approximated for all chemical mixtures using the coefficients for $^{12}$C.

\section{Electron captures fits}\label{apx_sec:ec_fits}

The analytic fits were made considering the tabulated electron capture rates from \cite{Suzuki_2016}, as reported by the NSCL group\footnote{\url{https://groups.nscl.msu.edu/charge_exchange/weakrates.html}} and plotted in Fig.~\ref{fig:ec_a56}. Below $T_9 = 2$, we observe these rates behave, to a very good degree, as $T$-independent. This motivated the use of a sixth degree polynomial in $x = \text{log}_{10}\rho$ for both $^{56}$Ni and $^{56}$Co electron capture rates $\lambda_{ec}$, such that
\begin{equation}
\text{log}_{10}\lambda_{ec}(x) = \sum^{6}_{j=0}a_{j}x^{j}.
\end{equation}
Since the lowest density entry in the table corresponds to $10^{5}$ g cm$^{-3}$, and we intended for our fit to provide a smooth transition towards $\lambda_{ec}\to 0$ at lower densities, artificial values were introduced in the fitted array in order to guarantee the polynomial remained strictly increasing in the range $[10^{0},10^{5}]$ g cm$^{-3}$.

For $^{56}$Ni we obtained:
\begin{align}
\text{log}_{10}\lambda_{ec}(x) = & -112.893743 + 57.5653413 x\nonumber\\
& - 9.4429386 x^2 - 0.04178759 x^3\nonumber\\
& + 0.16708368 x^4 - 0.01639904 x^5\nonumber\\
& + 4.953830726\times 10^{-3} x^6 .
\end{align}
while for $^{56}$Co we obtained:
\begin{align}
\text{log}_{10}\lambda_{ec}(x) = & -225.507327 + 173.403715 x\nonumber\\
& - 56.492487 x^2 + 9.673906 x^3\nonumber\\
& - 0.918252 x^4 + 4.614900 \times 10^{-2} x^5\nonumber\\
& -9.629495\times 10^{-4} x^6 .
\end{align}
%


\bsp	
\label{lastpage}
\end{document}